\documentclass[10pt,aps,prl,twocolumn,superscriptaddress,reprint,showpacs,balance,notitlepage]{revtex4-1}

\usepackage{amssymb,amsmath,mathtools}
\usepackage{graphicx}
\usepackage{color}
\usepackage{hyperref,tikz,pgfplots}
\usepackage{listings}
\usepackage{orcidlink}
\usepackage{balance}
\usepackage{soul}
\graphicspath{{figures/}}

\makeatletter
\newcommand*{\balancecolsandclearpage}{%
  \close@column@grid
  \clearpage
  \twocolumngrid
}
\makeatother

\begin{document}

\definecolor{dkgreen}{rgb}{0,0.6,0}
\definecolor{gray}{rgb}{0.5,0.5,0.5}
\definecolor{mauve}{rgb}{0.58,0,0.82}

\newcommand{\todo}[1]{\textbf{\textcolor{red}{#1}}}

\definecolor{mygreen}{rgb}{0.14, 0.84, 0.11}

\lstset{frame=tb,
  	language=Matlab,
  	aboveskip=3mm,
  	belowskip=3mm,
  	showstringspaces=false,
  	columns=flexible,
  	basicstyle={\small\ttfamily},
  	numbers=none,
  	numberstyle=\tiny\color{gray},
 	keywordstyle=\color{blue},
	commentstyle=\color{dkgreen},
  	stringstyle=\color{mauve},
  	breaklines=true,
  	breakatwhitespace=true
  	tabsize=3
}

\title{Passive viscous flow selection via fluid-induced buckling}

\author{Hemanshul Garg\,\orcidlink{0000-0002-0252-5877}}
\thanks{These authors are listed in alphabetical order.}
\affiliation{Department of Mechanical and Production Engineering, {\AA}rhus University, Inge Lehmanns Gade 10, 8000 {\AA}rhus C, Denmark}

\author{Pier Giuseppe Ledda\,\orcidlink{0000-0003-4435-8613}}
\thanks{These authors are listed in alphabetical order.}
\affiliation{
Department of Civil, Environmental Engineering and Architecture, University of Cagliari, Via Marengo 2, 09123 Cagliari, Italy
}%

\author{Jon Skov Pedersen}
\thanks{These authors are listed in alphabetical order.}
\affiliation{Department of Mechanical and Production Engineering, {\AA}rhus University, Inge Lehmanns Gade 10, 8000 {\AA}rhus C, Denmark}

\author{Matteo Pezzulla\,\orcidlink{0000-0002-3165-8011}}
\email{matt@mpe.au.dk}
\affiliation{Department of Mechanical and Production Engineering, {\AA}rhus University, Inge Lehmanns Gade 10, 8000 {\AA}rhus C, Denmark}

\date{\today}

\begin{abstract}
We study the buckling of a clamped beam immersed in a creeping flow within a rectangular channel. Via a combination of precision experiments, simulations, and theoretical modeling, we show how the instability depends on a pressure feedback mechanism and rationalize it in terms of dimensionless parameters. As the beam can bend until touching the wall above a critical flow rate, we finally demonstrate how the system can be used as a tunable passive flow selector, effectively redirecting the flow within a designed hydraulic circuit.
\end{abstract}

\maketitle
The efficient redistribution and control of flow is essential in many biological and engineered structures, from our cardiovascular system to plants and soft robots \cite{verzicco2022electro,Aylmore1984,Wehner2016}. For instance, plants majestically control and distribute the fluid flow within their lymphatic systems, without the need of any cerebral tissue and external actuation \cite{Aylmore1984}. Inspired by the biological world, microfluidic devices have been engineered with passive valves to enhance a variety of functions, ranging from cell manipulation to mixing and reacting devices \cite{Stone2004}, giving rise to the field of soft hydraulics, where the compliance of valves and channels is exploited to achieve new functionalities \cite{Leslie2009,Holmes2013,Reis2015,Christov2021,Louf2020}. Research efforts on passive control strategies have for example led to the design of fluidic diodes \cite{Leslie2009} and flow regulators \cite{Holmes2013,Gomez2017}. These applications have benefited from classical studies within the field of fluid-structure interactions \cite{Paidoussis1973,Grigorev1979}, but have also called for a better understanding of the behavior of flexible structures in fluidic channels, motivating studies on fixed \cite{wexler2013,Gosselin2014} and moving \cite{duroure2019,Chakrabarti2020,Cappello2022} fibers, and on flexible sheets \cite{Schouveiler2013,mahravan2023}. Within the field of soft hydraulics, the buckling of a clamped elastic fiber in a fluidic channel promises to be a good candidate to design tunable passive flow selectors, which would enrich the current ensemble of passive valves and the understanding of instabilities of flexible elements within microfluidic devices.

\begin{figure*} 
    \centering
    \includegraphics[width=1\textwidth]{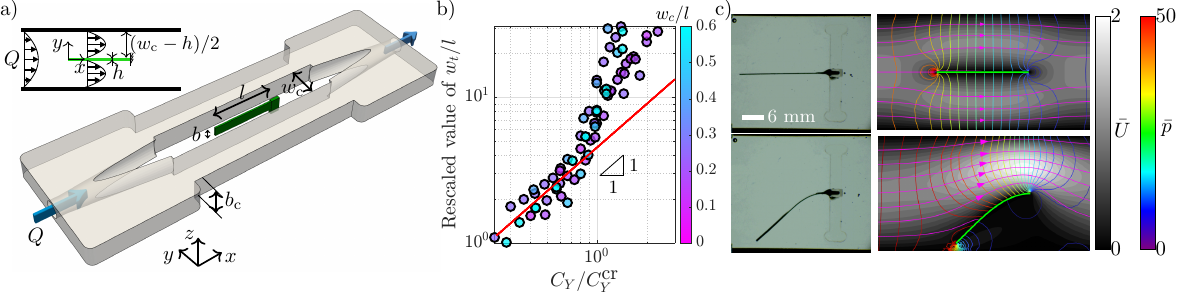} 
    \caption{(a) Schematic of the clamped beam inside a narrow channel. The Poiseuille flow is represented in the detailed sketch in the top left corner. (b) Experimental measurements. Tip displacement $w_t/l$ rescaled with the first observed value versus the critical Cauchy number normalized with the inferred threshold value $C_Y/C_Y^\textup{cr}$ to highlight the linear regime of the tip displacement, as denoted by the red solid line, for several $w_c/l$ according to the color bar. (c) Snapshots of straight and buckled beams (left: experiments; right: simulations). Color bars denote the dimensionless velocity $\bar{U}$ and $\bar{p}$ in simulations.}
    \label{fig:fig1}
\end{figure*}

In this Letter, we combine precision experiments with fluid-structure simulations and theoretical developments to reveal how Stokes flows induce beam buckling in fluidic channels. Our experiments demonstrate that, above a critical fluid load, the beam undergoes a buckling instability, thus bending to one side of the channel and behaving as a passive flow selector (Fig.~\ref{fig:fig1}).  As the problem naturally involves several geometric and material parameters pertaining to the beam, the channel, and the fluid, we carry out a dimensional analysis that untangles the physics of the problem and allows for a systematic exploration of the parameter space. In parallel, we perform two-dimensional (2D) and three-dimensional (3D) simulations of elastic beams immersed in a Stokes flow, and develop a theoretical model to rationalize our findings. We finally demonstrate that our results can inform the design of a tunable passive flow selector via a combination of experiments and 3D simulations, and that the geometry of the selector can be tailored to finely tune the flow rates at the outlets.

In our experiments, we fabricated thin elastomeric beams of two different materials: silicone-based vinylpolysiloxane (VPS) 32 (Zhermack) and PET (Mylar\textregistered, DuPont Teijin Films). For the former, we coated a smooth acrylic plate with the polymeric mixture and used a thin-film applicator (Futt, KTQ-II) to obtain layers with predefined and homogeneous thicknesses $h\in[0.25,0.6]$~mm. We then cut beams with height $b=3$~mm and length $l\in[7,30]$~mm. For PET beams, we used Mylar sheets with thicknesses $h\in[0.05,0.25]$~mm and cut beams with height $b\in[1,5.89]$~mm and length $l\in[6.9,40]$~mm. By performing self-buckling tests \cite{Greenhill1881}, we measured the Young's modulus for the two materials, resulting in $E=1.1\pm0.1$~MPa for VPS and $E=5.1\pm0.1$~GPa for PET (See Supplemental Material for further detail \footnote{See Supplemental Material for a detailed derivation, which includes Refs. \cite{white2006viscous,boussinesq1868memoire,Lee2016,Greenhill1881,ferreira2021hydrodynamic,Koiter1945,duprat2016fluid,Winkler1867}.}). A clamp holder secures the beam within a 3D printed channel with width $w_c\in[5,30]$~mm and height $b_c=6.5$~mm, as depicted in Fig.~\ref{fig:fig1} (a), where a flow rate~$Q\in[0.1,100]$ mL$/$min of silicone oil (dynamic viscosity $\mu=1\pm0.1$ Pa s) is driven by a syringe pump (Harvard Apparatus PHD Ultra 70-3007). A scientific camera (Basler Ace acA4096-40uc) is positioned above the channel to record the beam deformation, extracted via a custom MATLAB image processing code. 
Within this range of parameters in our experiments, the maximum Reynolds based on the hydraulic diameter of the channel was $0.58$, so that fluid inertia was negligible \cite{Note1}. In a typical experiment, we impose a flow rate~$Q$, achieved after a short preset ramp, and perform subsequent runs at increasing values of~$Q$ while recording the beam deformation. In each experiment, the tip displacement of the beam increases monotonically to a steady and constant value, following a short transient~\cite{Note1}. Above a critical flow rate, which depends on the geometrical and material parameters of the system, the beam deforms from the initial straight shape in Fig.~\ref{fig:fig1}~(c, top) to the bent configuration represented in Fig.~\ref{fig:fig1}~(c, bottom).

By means of dimensional analysis, we introduce the Cauchy number $C_Y=\mu U_\textup{max}l^2/E\hat{I}$, where $U_\textup{max}$ is the maximum velocity as given by the 3D Poiseuille flow at the inlet, and $\hat{I}=h^3/12$ is the moment of inertia per unit width of the beam \cite{Note1}. The dimensionless number $C_Y$ represents the ratio between the fluid ($\sim\mu U_\textup{max}/l$) and elastic stresses ($\sim E\hat{I}/l^3$), thereby combining some geometrical parameters of the system with the material parameters of the beam and the fluid \cite{Gosselin2010}. By assuming $w_c \simeq w_c - h=w_c^*$, we can further reduce the number of parameters at play. Therefore, a critical flow rate corresponds to a critical Cauchy number $C_Y^\textup{cr}${, beyond which the beam diverges from the initial straight shape}, which depends only on the remaining geometrical parameters $w_c^*/l$ and $b/b_c$. To quantitatively define the critical Cauchy number $C_Y^\textup{cr}$, we analyze the steady-state (maximum) tip displacement $w_t/l$ of the beam as a function of~$C_Y$, as shown in Fig.~\ref{fig:fig1}~(b). The tip displacement, rescaled by the first observable value, presents a linear growth with the Cauchy number, rescaled by the critical one~$C_Y^\textup{cr}$ (as determined from the experimental data), then followed by a sudden superlinear regime, similarly to the Euler buckling of thin beams with small imperfections~\cite{timoshenko1976}. As a protocol, we define the critical Cauchy number~$C_Y^\textup{cr}$ as the lowest Cauchy number corresponding to a relative variation of $w_t/l$ of $5\%$ from the linear trend~\cite{Note1}.

\begin{figure}[t]
\centering
\includegraphics[scale=0.825]{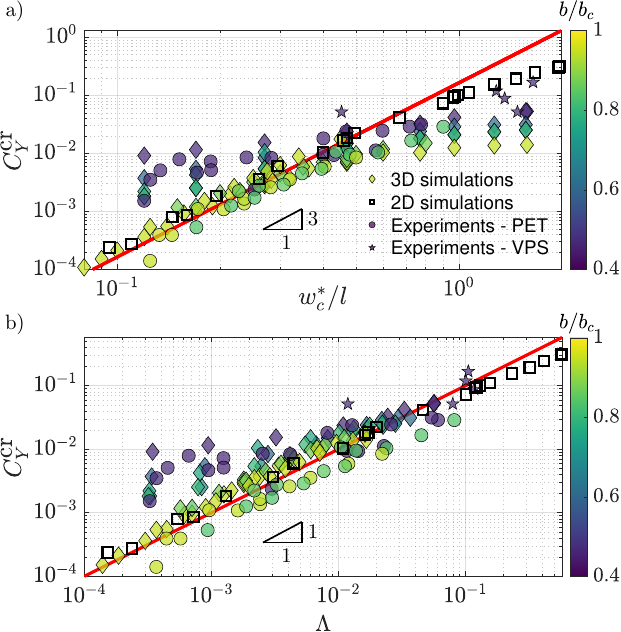}

    \caption{Fluid-induced buckling of clamped beams. (a) Critical Cauchy number $C_Y^\textup{cr}$ as a function of $w_c^*/l$ for several $b/b_c$, as shown by the color bar, from experiments and simulations. The red solid line denotes the cubic law from the analytical solution under the assumption of 2D flow. (b) Critical Cauchy number $C_Y^\textup{cr}$ as a function of the dimensionless geometric parameter $\Lambda$, with the red solid line denoting the analytical solution in Eq.~\eqref{eq:lambda} with three-dimensional and wall shear stress effects.}
    \label{fig:fig2}
\end{figure}

To improve our initial understanding of the experimental results, we perform 2D and 3D fluid-structure simulations by solving the dimensionless Stokes equations coupled with the balance equations of Hookean solids undergoing small strains but large displacements, enforcing stress continuity at the fluid-solid interface~\cite{Note1}. Fig.~\ref{fig:fig2}~(a) shows the critical Cauchy number as a function of $w_c^*/l$ and $b/b_c$ as obtained from simulations and experiments. The slope of the red solid line denotes the cubic scaling $C_Y^\textup{cr}\sim(w_c^*/l)^3$ observed for $b/b_c\rightarrow1$ and $w_c^*/l<1$, that is a high-confinement regime{, as rationalized later}. Experiments and 3D simulations are in good agreement over a wide range of parameters, with 2D simulations replicating the behavior of the system for high confinement ratios.

To rationalize our experimental and numerical results, we first develop a 2D theoretical model. We assume a parabolic Poiseuille flow profile inside the channel, including the gaps between the beam and the walls (Fig.~\ref{fig:fig1}~(a)), meaning that the pressure~$p$ does not vary along the cross-stream direction, denoted by $y$, when the beam is straight. This can also be seen from our 2D simulations depicted in Fig.~\ref{fig:fig1}~(c), where the flow rate~$Q$ splits into two flow rates $Q/2$ within the two gaps, above and below the beam. Within each gap~$H$, the streamwise-invariant Poiseuille flow is characterized by a pressure gradient $G=\partial p/\partial x={6 \mu Q}/{H^{3}}$, where $x$ is the streamwise coordinate such that $x=0$ at the free tip of the beam (Fig.~\ref{fig:fig1}~(a)). 
At the onset of buckling, the beam deflects with a vertical displacement $w(x) \ll w_c^*$, such that the gap of the upper (+) and lower (-) parts becomes $H(x)=w_c^*{/2} \mp w(x)$. Therefore, upon integration from the common pressure value at the leading edge of the beam ($x=0$), the pressure field becomes $p_{\pm}(x) \simeq - 6\mu Q x /\left(w_c^* / 2\mp w(x)\right)^3 +p(0)$, where we neglected the dependence of $H$ with $x$ while integrating along the beam, since $w(x) \ll w_c^*$. At a fixed downstream position, the transverse load per unit length due to the pressure difference between the two sides of the beam is expressed as a Taylor series for $w(x)/w_c^* \rightarrow 0$:
\begin{multline}
    q_y (x)\boldsymbol{e}_y=-p_{+}\left(\boldsymbol{e}_y\right)+p_{-}\left(\boldsymbol{e}_y\right) =  \\ \frac{576 \mu Q x}{w_c^{*3}} \frac{w(x)}{w_c^*}\boldsymbol{e}_y +\mathcal{O}(w^3(x)) \simeq \alpha x {w(x)}\boldsymbol{e}_y\,,
\end{multline}
where $\boldsymbol{e}_y$ is the unit basis vector along $y$ {and $\alpha$ is defined}. This force acts along the same direction of the displacement and represents a positive feedback due to beam deflection. This pressure imbalance can be appreciated by the pressure iso-contours in Fig.~\ref{fig:fig1}~(c).
Buckling instability occurs when the transversal pressure load, which increases with the deflection of the beam, overcomes the bending internal stresses of the beam:
\begin{equation}
    EI\frac{w_t}{l^4} \sim  \frac{ \mu Q b l}{w_c^{*3}}\left(\frac{w_t}{w_c^*}\right) \Rightarrow     C_Y^{\textup{cr}} \sim\left(\frac{w_c^*}{l}\right)^3,
\end{equation}
where we used the tip displacement~$w_t$ as a representative displacement and $Q \simeq (2/3) U_\textup{max} w_c^*$ \cite{Note1}. This theoretical prediction agrees with the cubic scaling found in Fig.~\ref{fig:fig2}~(a) for high confinement{, where we plotted it with the prefactor $0.1637$ derived via a quantitative linear stability analysis \cite{Note1}}.

\begin{figure}
\includegraphics[scale=1.08]{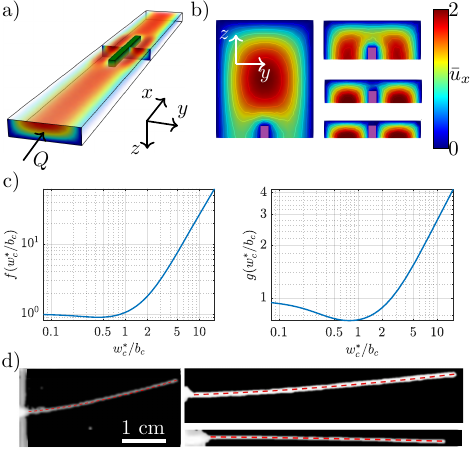}
\caption{(a) Three-dimensional structure of the flow in the channel for $w_c/l=0.7$. (b) Slices of the flow structure in the yz-plane for $b/b_c=0.1,0.3,0.7,0.9$. Color bars denote the dimensionless velocity component in the x-direction.
(c) Functions $f$ (left) and $g$ (right) versus $w_c^*/b_c$. (d) Experimental deformed shapes of the beam (white) for $w_c^*/l=0.47$, $b/b_c=0.46$ (left), $w_c^*/l=0.14$, $b/b_c=0.46$ (right, top), $w_c^*/l=0.16$, $b/b_c=0.46$ (right, bottom) overlaid with the theoretical approximation of the post-buckling beam deflection $w^{(0)}$ (red dashed lines).}
\label{fig:fig3}
\end{figure}

{However, as $w_c^*/l \rightarrow 1$, a progressive deviation from the cubic trend is observed, for both 2D and 3D settings. For 2D, the deviation from the cubic trend is due to the decreased pressure-feedback:} the compression load {per unit length, $q_x=2\tau$}, due to the wall shear stresses, $\tau=G{H}/{2} \simeq {8\mu U_\textup{max}}/{w_c^{*}}$, 
becomes more important as $w_c^*/l$ increases, since the pressure-driven feedback decreases {as $q_y \propto {w_c^*}^{-2}$, while wall shear stresses as $q_x \propto {w_c^*}^{-1}$.} 
{For 3D, this effect occurs for smaller $w_c^*/l$ with increasing $b/b_c$, since channels have a more slender cross-section, \emph{i.e.}, a larger $w_c^*/b_c$ (see Fig.~\ref{fig:fig3}~(b)). Indeed, for $w_c^*/l \sim 1$, a geometry with $l\gg b$ and $b/b_c \sim 1$ imply $w_c^* \gg b_c$, \emph{i.e.} a shallow channel. For the same maximum velocity, hydrodynamic forces increase for shallower channels \cite{Gomez2017}.}
A second improvement can be {thus} obtained by modeling the 3D effects due to the aspect ratio of the cross section of the channel~\cite{boussinesq1868memoire}, which have been neglected so far. Indeed, for $b/b_c\gtrapprox0.5$, two 3D Poiseuille profiles stand on the sides of the beam along the y-axis {(Fig.~\ref{fig:fig3}~(a,b))}.
By taking this 3D structure into consideration, we can calculate the 3D pressure gradient and wall shear stresses as 
\begin{equation}
    G=f\left(\frac{w_c^*}{b_c}\right)\frac{32\mu U_\textup{max}}{w_c^{*2}}\,, \ \ \tau=g\left(\frac{w_c^*}{b_c}\right)\frac{8 \mu U_\textup{max}}{w_c^{*}}\,, 
\end{equation}
where $f(w_c^*/b_c)$ and $g(w_c^*/b_c)$ are analytical functions of the aspect ratio of the {channel} cross section 
{(Fig.~\ref{fig:fig3}~(b,c)).} 

We proceed to obtain a quantitative prediction of the buckling threshold by means of linear beam theory. 
The Euler-Bernoulli beam equilibrium equation under the transversal load $q_y$, proportional to the beam displacement $w(x)$,  {and the constant compressive load $q_x$} reads
$ E \hat{I} w^{\prime \prime \prime \prime}(x) {+q_x\left({x}{{w}^\prime}({x})\right)^\prime- q_y(x)}=0$ \cite{Greenhill1881,timoshenko1976}. 
Upon non-dimensionalization with the {beam length}
, we obtain
\begin{multline}
    \bar{w}^{\prime \prime \prime \prime}(\bar{x})+16C_Y g(w^*_c/b_c) \frac{l}{w_c^*}\left(\bar{x}\bar{w}'(\bar{x})\right)' - \\ -384 C_Y f(w^*_c/b_c)\left(\frac{l}{w_c^*}\right)^3 \bar{x} \bar{w}(\bar{x})=0\,,
    \label{eq:buckling_tot}
\end{multline}
completed with the classical free-edge ($\bar{w}''(0)=\bar{w}'''(0)=0$) and clamp boundary conditions ($\bar{w}(1)=\bar{w}'(1)=0$), where bars denote non-dimensional variables. A standard linear stability analysis looks for non-trivial solutions of this homogeneous problem to evaluate the critical value of $C_Y$. 
For an imposed tip displacement $w_t$, a straightforward guess of the beam displacement {that satisfies boundary conditions}, neglecting the distributed nature of the pressure load, is the third-order polynomial
$ w^{(0)}(x)=\frac{w_t}{2}\left( 2-3\left(\frac{x}{l}\right)+\left(\frac{x}{l}\right)^3\right)${, which aligns with experimental deformed shapes (Fig.~\ref{fig:fig3}~(d)).}
An approximation of the instability threshold is thus obtained by injecting the post-buckling beam deflection $w^{(0)}(x)$ as guess for $w(x)$, \textit{i.e.} solving
  $\bar{w}^{\prime\prime\prime\prime}(\bar{x}) +\bar{\beta}\left(\bar{x}{\bar{w}^{(0)\prime}}(\bar{x})\right)^\prime - \bar{\alpha} \bar{x} \bar{w}^{(0)}(\bar{x}) = 0$ with the same boundary conditions and imposing the same free-edge displacement $w(0)=w_t$ ($\bar{\beta}$ is defined from \eqref{eq:buckling_tot}). We obtain the compatibility condition $7\bar{\alpha}/480+\bar{\beta}/8=1$, leading to 
\begin{equation}
    C_Y^{\textup{cr}}={\frac{1}{16}\frac{w_c^*}{l}}\left({\frac{1}{8}g\left(\frac{w_c^*}{b_c}\right)+\frac{7}{20}\left(\frac{l}{w_c^*}\right)^2f\left(\frac{w_c^*}{b_c}\right)}\right)^{-1}\!\!\!\eqqcolon\!\Lambda\,,
\label{eq:lambda}
\end{equation}
where we define the dimensionless geometric function~$\Lambda$, depending on the 3D geometry of the system. 
{For $f=g=1$ and ${w_c^*}/l\ll1$, this reduces to the 2D pressure-driven case, $C_Y^{\textup{cr}} \simeq 0.1786 ({w_c^*}/l)^3$, very close to the theoretical exact prefactor \cite{Note1}.}
In Fig.~\ref{fig:fig2}~(b), we plot our numerical and experimental results as a function of~$\Lambda$, showing an overall collapse of the data as predicted by Eq.~\eqref{eq:lambda} (red solid line), without any fitting parameters. The {remaining} deviation{s} from theory 
can be attributed to the gap between the channel and the beam along the z axis
{when $1-b/b_c \gg w_c^*/l$}
, which weakens the 2D pressure-driven feedback mechanism as the beam deflects {since the flow mostly escapes through the gaps along the z axis (Fig.~\ref{fig:fig3}~(b))}. 

\begin{figure}[b] 
\vspace{-0.5cm}
    \centering

\includegraphics[scale=0.98]{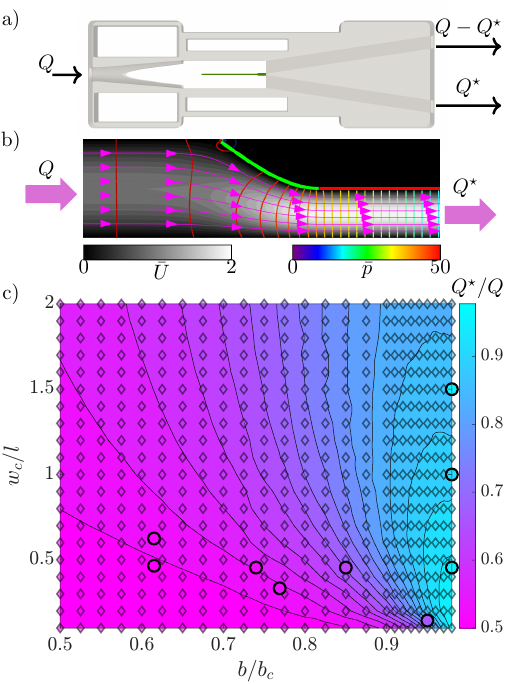}
    
    \caption{Buckled beams as passive flow selectors. (a) 3D printed channel used for the experiments. (b) Streamlines and contour pressure lines from 2D simulations for a fully buckled beam touching the wall. The flow rate $Q$ is redirected to the bottom channel with flow rate $Q^\star$ ($Q=Q^\star$ in this case). Color bar denotes the dimensionless velocity $\bar{U}$ and $\bar{p}$ in simulations. (c) Contour map of $Q^\star/Q$ in the ($b/b_c \in [0.5 , 0.98]$, $w_c/l \in [0.1, 2]$) plane. Diamonds and contour lines denote 3D simulations while experiments are represented by circles.}
    \label{fig:fig4}
\end{figure}

To demonstrate that our system can be used to design a passive flow selector, we perform experiments and simulations in a channel where a beam is placed right upstream of a bifurcation, as depicted in the 3D printed channel in Fig.~\ref{fig:fig4}~(a). A flow rate $Q$ above the critical one was imposed via the syringe pump, and the flows from the two different outlets were collected in two different cylindrical containers. By tracking the height of the fluids over time, we were able to measure the flow rates $Q^\star$ and $Q-Q^\star$ in the outlets. A simplified 2D case is depicted in Fig.~\ref{fig:fig4}~(b). If the beam (in green) deflects to one side until it touches the wall, 2D simulations show that the flow is redirected to the opposite side, with the flow at the outlet equal to the flow rate at the inlet~$Q^\star=Q$. However, as experiments are inherently three-dimensional, we expect~$Q^\star\neq Q$ in general, as the fluid can move above and below the beam through its lateral ends, since $b/b_c<1$.  
Therefore, we perform 3D simulations with a rigid beam touching the wall, by taking advantage of the post-buckling shape derived above, to construct a phase map of the relative flow rate $Q^\star/Q$ as a function of $b/b_c$ and $w_c/l$, in the case where the beam touches the wall. Fig.~\ref{fig:fig4}~(c) depicts this phase map, where diamonds and contour lines denote 3D simulations, while experiments are represented by circles. The map shows that $Q^\star/Q$ can be finely tuned by a careful selection of the geometrical parameters $b/b_c$ and $w_c/l$, while the 2D case, where $Q^\star=Q$, can be recovered for $b/b_c\rightarrow1$ and $w_c/l<2$.

In summary, we have demonstrated how a clamped beam in a channel can undergo a buckling instability, which can be harnessed to design a tunable passive flow selector. We have developed a 3D theoretical model that reveals a nontrivial pressure feedback, which governs the high-confinement regime, and successfully combines the relevant material and geometrical parameters of the system. Albeit the direction of buckling in our experiments is undetermined a priori, as it depends on imperfections~\cite{Note1}, we anticipate that it can be encoded in the system by seeding precise defects or designing a bilayer beam that realizes a natural curvature due to variations in temperature \cite{Morimoto2015} or pH \cite{Jin2018}. This system may find application in microfluidic systems such as cell-sorting \cite{Shields2015}, or provide a simple solution in applications where the flow has to be redirected passively to specific appendices, such as in soft robotics \cite{Wehner2016}. Lastly, we envision this system to be employed for the  indirect measurements of elastic properties of small and soft fibers \cite{duprat2016fluid,Cappello2022,Liu2024}, where standard mechanical tests fail, which we hope the current study will motivate.

\begin{acknowledgments}
This work was supported by a research grant (VIL50135) from VILLUM FONDEN. M.P. acknowledges also the support from the Thomas B. Thriges Fond. H.G., P.G.L. and M.P. wrote the
manuscript. M.P. conceived the project, supervised the research, and performed the numerical simulations with inputs from P.G.L. H.G. conducted the experiments. J.S.P. conducted the preliminary numerical explorations and designed the experimental setup. P.G.L. developed the theoretical models. 
\end{acknowledgments}

\bibliography{biblio}

\balancecolsandclearpage
\newpage

\onecolumngrid

\begin{center}
\textbf{\large Supporting information}
\end{center}

\setcounter{equation}{0}
\setcounter{figure}{0}
\setcounter{table}{0}
\setcounter{page}{1}
\makeatletter
\renewcommand{\theequation}{S\arabic{equation}}
\renewcommand{\thefigure}{S\arabic{figure}}

\section{Experimental details}

\subsection{Experimental apparatus}
\begin{figure}[h]
\begin{tikzpicture}[scale=1]

			\node[inner sep=0pt] at (0.8,2.725-0.4)
	{\includegraphics[scale=0.12]{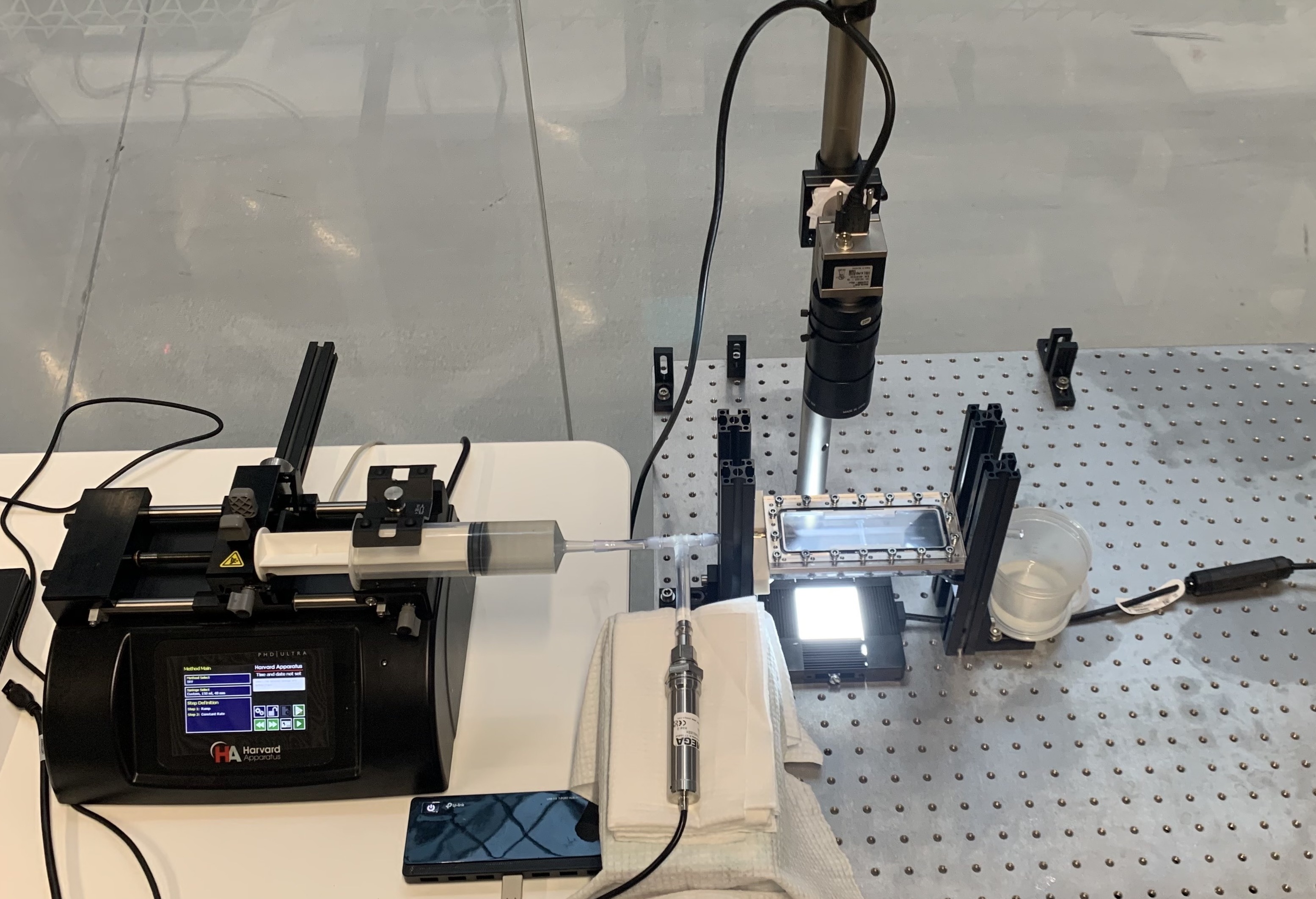}};

 			\node[inner sep=0pt] at (-3.65,5.98)
	{\includegraphics[scale=0.2]{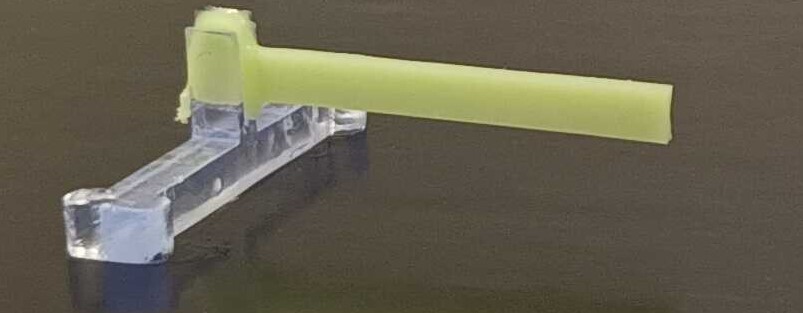}};
	
     \draw[->,white,thick] (1.65,7.465) -- (1.65,7.68);

	      \node[inner sep=0pt,white] at (-3.05,2.55)
	{\textbf{Syringe pump}};
	      \node[inner sep=0pt,white] at (1.45,4.55)
	{\textbf{Camera}};	
	      \node[inner sep=0pt,black] at (2.45,-1.55)
	{\textbf{Pressure sensor}};	
 	      \node[inner sep=0pt,white] at (2.95,-0.45)
	{\textbf{Backlight}};	
  	      \node[inner sep=0pt,white] at (3.9,3)
	{\textbf{Channel}};
     \draw[->,white,very thick] (3.9,2.8) -- (3,2);	
      
\end{tikzpicture}
\caption{Photo of the experimental apparatus comprising the syringe pump, the pressure sensor, the scientific camera with a backlight, and the 3D printed channel hosted within the aluminum frame. A detail of a VPS beam and its clamp is depicted in the top left corner.}
\label{fig:setup}
\end{figure}

A schematic of the complete experimental setup is shown in Figure \ref{fig:setup}. The channel consists of several components: the main frame is made of aluminum and covered by transparent acrylic sheets from top and bottom. A backlight  (Edmund Optics AI Side-Fired Backlight, $2" \times 2"$, White) is placed below the channel to help with image processing. A scientific camera (Basler Ace acA4096-40uc USB3 with a color zoom lens 13-130 mm) is mounted at the top of the channel and employed to record and capture the beam deformation. A family of 3D printed channels with different widths, placed within the main aluminum frame, allows for varying the channel width. Within each 3D printed geometry, the channel width is gradually increased from the inlet diameter to the desired width. We considered channel widths in the range $w_c \in [0.5,3]$ cm and a fixed channel height $b_c = 0.65$ cm. 

\begin{figure}
    \centering
    \includegraphics[scale=0.46]{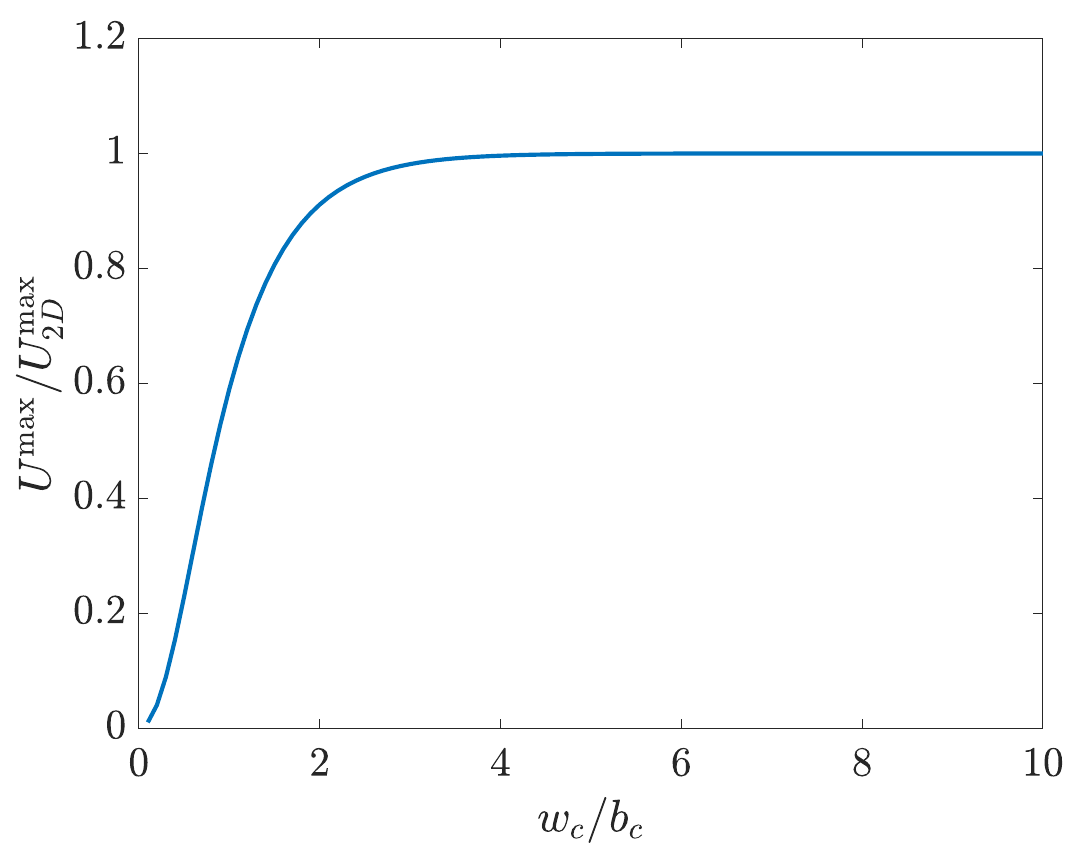}
    \caption{{Dependence of $ U_\textup{max}/U_\textup{max}^{2D}$ on the aspect ratio $w_c/b_c$. }}
    \label{fig:Umax}
\end{figure}

Silicone oil (Sigma-Aldrich, kinematic viscosity, $\nu=1000$ cSt and density $\rho= 970$ kg m$^{-3}$) is used as the working fluid and the discharge flow, $Q$, is manipulated by a syringe pump (Harvard Apparatus PHD ULTRA Syringe Pump 70-3007). We employed syringes with varying capacity (ACONDE 20-150 mL plastic syringe) and all experiments were performed with $0.1 \le Q \le 100$ mL min$^{-1}$. Flexible PVC tubes are used to connect the syringe and the pressure sensor (OMEGA PXM409-170HGUSBH) to the channel. The Reynolds number ($\textup{Re}_{D_h}=D_h U_\textup{max}/\nu$), based on the hydraulic diameter $D_h=2b_cw_c/(b_c+w_c)$ \cite{white2006viscous} and the fully developed maximum flow velocity $U_\textup{max}$, ranges from $10^{-4}$ to $0.58$. 
To evaluate the maximum flow velocity in the fully developed region, we used the 3D analytical solution of a Poiseuille flow through channels of rectangular cross-section \cite{boussinesq1868memoire} :

\begin{equation}\label{eq:velocity}
u(z{^*},y{^*}) = \frac{G}{2\mu}z{^*}(b_c-z{^*}) - \frac{4Gb_c^2}{\mu\pi^3}\sum_{n=1}^{\infty}\frac{1}{(2n-1)^3}\frac{\sinh(\alpha_n y{^*}) + \sinh[\alpha_n (w_c-y{^*})]}{\sinh(\alpha_n w_c)}\sin(\alpha_n z{^*}), \alpha_n = \frac{(2n-1)\pi}{b_c}\,, 
\end{equation}

\begin{equation}\label{eq:discharge}
Q = \frac{Gb_c^3w_c}{12\mu} - \frac{16Gb_c^4}{\pi^5\mu}\sum_{n=1}^{\infty}\frac{1}{(2n-1)^5}\frac{\cosh(\alpha_nw_c) - 1}{\sinh(\alpha_n w_c)}\,,
\end{equation}  
where $u(z{^*},y{^*})$ represents the velocity as a function of the coordinates $z{^*}$ and $y{^*}$ that run along the height and the width of the channel {($y^*=y+w_c/2,z^*=z+b_c/2)$}, respectively, and $Q$ represents the discharge flow rate. The constant pressure gradient is denoted as $G = -dp/dx$, while $\mu=\rho\nu$ is the dynamic viscosity. Combining Eq. \eqref{eq:velocity} with Eq. \eqref{eq:discharge}, we can express the maximum velocity as a function of the flow rate and employ it to determine the corresponding Cauchy number $C_Y = \mu U_\textup{max} l^2/(E \hat{I})$, where $\hat{I}= I/b = h^3/12$ is the second moment of inertia per unit width of the beam and $l$ is the length of the beam{, and $U_\textup{max}=u(b_c/2,w_c/2)$ reads:
\begin{equation}\label{eq:velocitymax}
U_\textup{max} = \frac{Gb_c^2}{8\mu}\left(1-\frac{32}{\pi^3}\sum_{n=1}^{\infty}\frac{1}{(2n-1)^3}\frac{\sin\left(\frac{(2n-1)\pi}{2}\right)}{\cosh\left(\frac{(2n-1)\pi}{2} (w_c/b_c)\right)}\right)\,. 
\end{equation}
Note that term $\frac{Gb_c^2}{8\mu}$ is the maximum velocity $U_\textup{max}^{2D}$ in the 2D case, i.e. as $w_c/b_c \rightarrow \infty$. The dependence of $ U_\textup{max}/U_\textup{max}^{2D}$ on the aspect ratio $w_c/b_c$ is reported in Figure \ref{fig:Umax}. For constant values of $G,\,b_c$ and $\mu$, the maximum velocity within the channel increases as the transversal length $w_c$ increases, until it saturates to the 2D value at $w_c/b_c\sim 5$. In tighter channels, the maximum velocity decreases for the same pressure gradient $G$. Therefore, to maintain the same maximum velocity as $w_c/b_c$ increases, the pressure gradient must also increase.}

\subsection{Beam fabrication and characterization}

We consider two different materials to fabricate the beams: VPS-32 (vinyl-polysiloxane, Zhermack) and PET (Mylar\textregistered, DuPont Teijin Films). VPS beams are prepared by mixing the bulk and curing agents at 1:1 mass ratio in a centrifugal mixer (Thinky Mixer ARE-250CE), at 1200 rpm for 20 seconds \cite{Lee2016}. The resulting fluid is poured on acrylic plates where an adjustable thin film coating applicator (Futt, KTQ-II) is used to achieve a specific thickness. Curing takes approximately $20$ min at room temperature. Mylar beams are instead prepared by cutting the desired shape out of $75$ $\mu$m, $100$ $\mu$m and $250$ $\mu$m thin sheets. All beams have a ratio $b/l<0.12$ so that they are well within the beam regime and far from the plate behavior. The densities for both materials are determined by measuring the mass of plate-like samples of known geometry with a precision scale (Kern, ABS220-4N). We found $\rho=1160$ kg$/$m$^3$ for VPS and $\rho=1416$ kg$/$m$^3$ for Mylar.

The Young's modulus, $E$, for both materials is determined by the self-buckling test \cite{Greenhill1881}. A beam with specified thickness $h$ and width $b$ is clamped vertically and its length is increased (by pushing it through the clamp) until buckling is observed. Then, the Young's modulus is estimated via the formula $l_\textup{max}=(7.8373 E\hat{I}/\rho g h)^{1/3}$ \cite{Greenhill1881}. By repeating the self-buckling experiment for different beam geometries, we found that $E = 1.1\pm$0.1 MPa for VPS and that $E = 5.1\pm$0.1 GPa for Mylar.

\subsection{Experimental procedure}

In each experimental run, the beam is secured to a 3D-printed detachable holder using VPS, allowing easy fixation within the channel. The beam length is adjusted so that its tip is positioned within the fully developed region of the fluid flow. This is ensured by computing the entrance length for a specific channel and flow rate as documented in \cite{ferreira2021hydrodynamic}.


At the beginning of each experiment, the channel is fully filled with silicone oil. We minimize the presence of bubbles by flushing the channel at a low discharge flow rate that does not induce buckling. When the channel is ready and the camera recording, the syringe pump is started and the flow rate is increased from $0$ to the desired value via a $15$ s ramp to achieve the desired steady-state flow rate while minimizing inertial effects so that, if the experiments are run with a longer ramp, no changes in the critical Cauchy number are observed. The steady-state flow rate is then imposed and the deformation of the beam is recorded. Each video is then processed via a custom MATLAB script to extract the deformed shape of the beam over time, for different experimental parameters.

\begin{figure}[!h]
\begin{tikzpicture}[scale=1]
			\node[inner sep=0pt] at (0.8,2.725-0.4)
	{\includegraphics[scale=0.31]{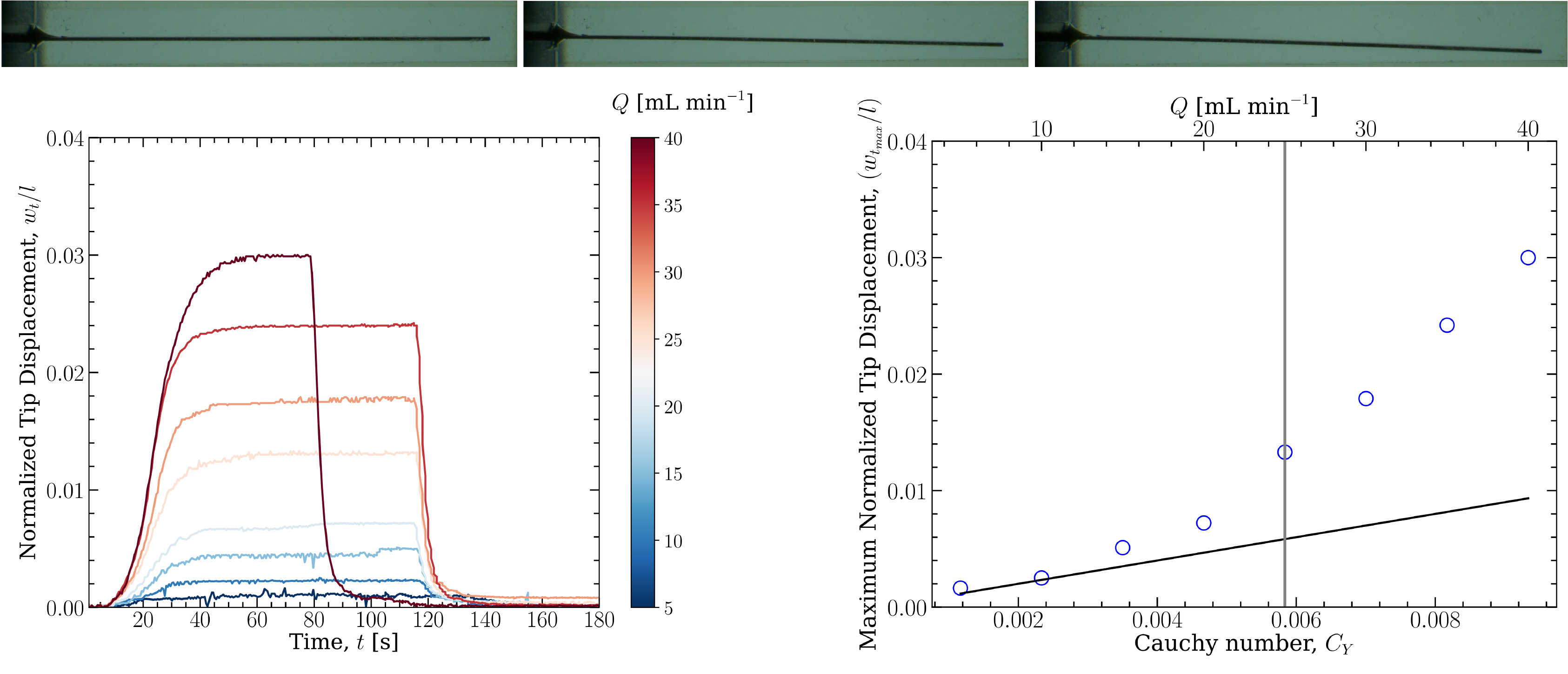}};

 \node[inner sep=0pt,black] at (-7.95,6.55)
	{a)};
  \node[inner sep=0pt,black] at (-5.05,6.3)
	{\text{$Q$ = 10 mL min$^{-1}$}}; 
  \node[inner sep=0pt,black] at (0.75,6.3)
	{\text{$Q$ = 25 mL min$^{-1}$}};
   \node[inner sep=0pt,black] at (6.55,6.3)
	{\text{$Q$ = 40 mL min$^{-1}$}};
  \node[inner sep=0pt,black] at (-7.95,5.05)
	{b)};
   \node[inner sep=0pt,black] at (1.1,5.05)
	{c)};
      
\end{tikzpicture}
\caption{Beam with $h/l = 0.0066$ in a channel with $w_c/l=0.13$ and $b/b_c = 0.46$. (a) Snapshot of the beam deformation at $Q=10$, $Q=25$, and $Q=40$ mL min$^{-1}$, respectively. (b) Dimensionless tip displacement $w_t/l$ versus time, for different flow rates, and (c) maximum dimensionless tip displacement ${w_t}_\textup{max}/l$ versus Cauchy number as well as flow rate. The maximum tip displacement would start to saturate for larger flow rates.}
\label{fig:beam34}
\end{figure}

\begin{figure}[!h]
\begin{tikzpicture}[scale=1]
			\node[inner sep=0pt] at (0.8,2.725-0.4)
	{\includegraphics[scale=0.4]{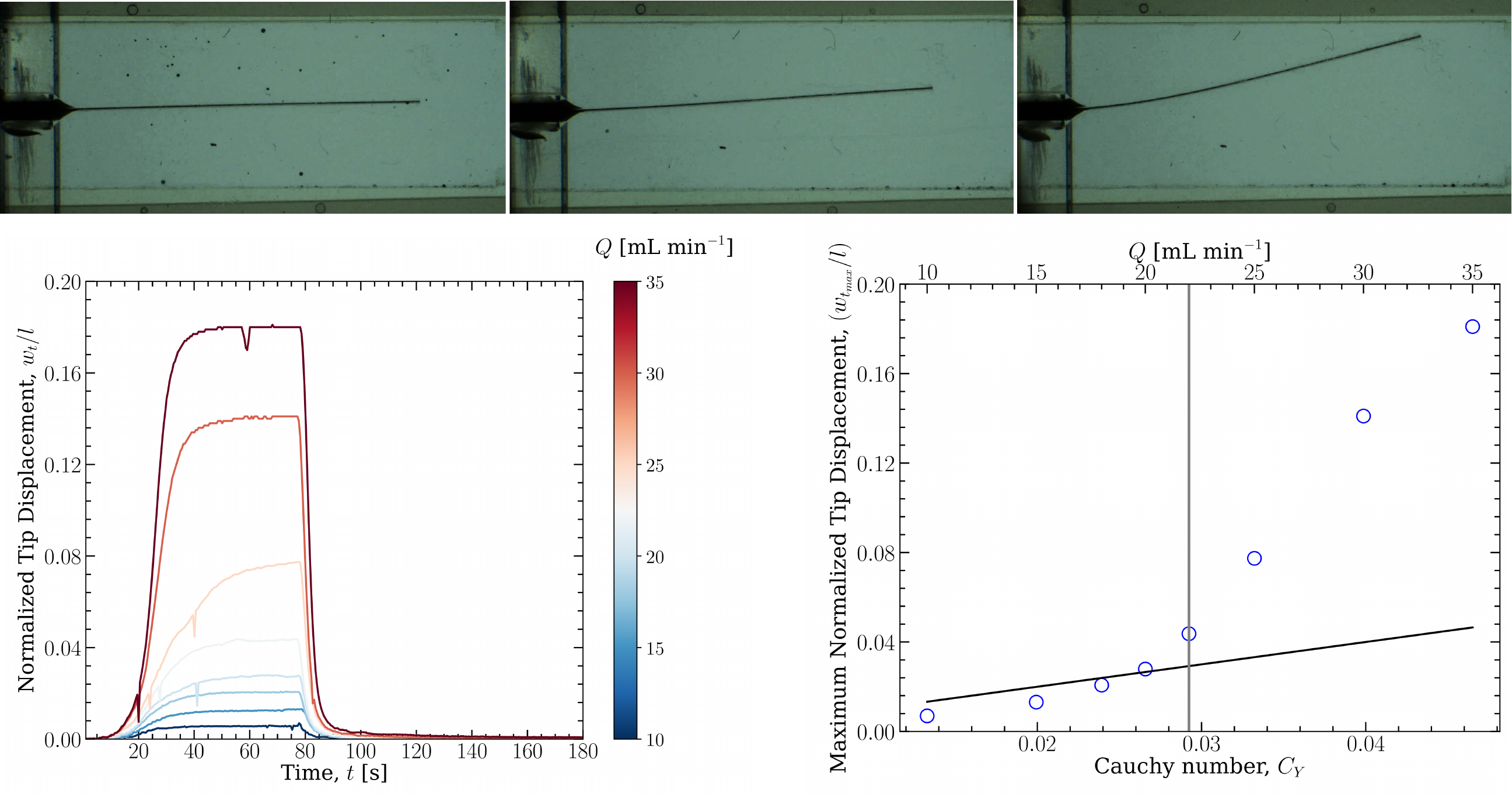}};
   \node[inner sep=0pt,black] at (-7.95,7.55)
	{a)};

   \node[inner sep=0pt,black] at (-5.05,7.3)
	{\text{$Q$ = 10 mL min$^{-1}$}}; 
  \node[inner sep=0pt,black] at (0.75,7.3)
	{\text{$Q$ = 22 mL min$^{-1}$}};
   \node[inner sep=0pt,black] at (6.55,7.3)
	{\text{$Q$ = 35 mL min$^{-1}$}};
  \node[inner sep=0pt,black] at (-7.95,4.05)
	{b)};
   \node[inner sep=0pt,black] at (1.05,4.05)
	{c)};    
\end{tikzpicture}
\caption{Beam with $h/l = 0.0035$ in a channel with $w_c/l=0.47$ and $b/b_c = 0.46$. (a) Snapshot of the beam deformation at $Q=10$, $Q=22$, and $Q=35$ mL min$^{-1}$, respectively. (b) Dimensionless tip displacement $w_t/l$ versus time, for different flow rates, and (c) maximum dimensionless tip displacement ${w_t}_\textup{max}/l$ versus Cauchy number as well as flow rate.
}
\label{fig:beam42}
\end{figure}

Two typical experiments are summarized in Figures \ref{fig:beam34} and \ref{fig:beam42}. For example, Figure \ref{fig:beam34} shows an experiment characterized by $w_c/l=0.13$, $h/l=0.0066$ and $b/b_c=0.46$. Snapshots at three different flow rates are depicted in Figure \ref{fig:beam34} (a), while the dimensionless tip displacement ($w_t/l$) is represented in Figure \ref{fig:beam34} (b). Additionally, Figure \ref{fig:beam34} (c) depicts the maximum dimensionless tip displacement (${w_t}_\textup{max}/l$) as a function of the Cauchy number. The vertical black line represents the buckling threshold discussed in the main text, denoting the end of the linear regime. As stated in the main, we define the critical Cauchy number~$C_Y^\textup{cr}$ as the lowest Cauchy number corresponding to a relative variation of the maximum value of $w_t/l$ of $5\%$ from the linear trend.
{More specifically, the linear trend is determined by performing a linear regression of the experimental values, starting with the first two (the two lowest values of $C_Y$). Then, if the next data point does not deviate by more than 5$\%$ with respect to the linear regression, the data point is added to perform a new linear regression with three data points. This procedure continues until the next experimental value deviates by more than 5$\%$ with respect to the value predicted by the linear regression for the same experimental value of $C_Y$, which is then identified as the critical Cauchy number $C_Y^\textup{cr}$; note that increasing the threshold up to 50$\%$ causes variations in the Cauchy number less than the marker size employed in the plots.}

Similarly, Figure \ref{fig:beam42} shows another experiment characterized by $w_c/l=0.47$, $h/l=0.0035$ and $b/b_c=0.46$. The estimation of the critical buckling Cauchy number is therefore affected by uncertainties in the experimental procedure and in the material and geometrical parameters. Consequently, for each experimental run, we propagate the uncertainties following the definition of the Cauchy number $C_Y = 12 \mu U_\textup{max} l^2/E h^3$, and determine error bars that result to be smaller than the symbol size in Figure 2 of the main text. Specifically, the Young's modulus is affected by an uncertainty in our measurement as outlined above, the geometrical parameters such as length and thickness are affected by the resolution of our camera as they are determined via image processing ($\Delta h/h\simeq3\%$, $\Delta l/l\simeq3\%$). Finally, the uncertainty in the viscosity is determined from the viscosity-temperature plot in the technical spreadsheet given by the producer ($\Delta \mu/\mu\simeq5\%$).


\section{Numerical details}

Numerical simulations are set up in COMSOL Multiphysics (v6.1) within the Fluid-Solid interaction package, for both 2D and 3D settings, where the dimensionless equations are solved. A time-dependent solver is employed to solve for the beam deformation and identify the buckling threshold, as outlined in the main text. 
{In this time-dependent setting, the threshold is identified as the value of $C_Y$ for which an exponential growth of the tip displacement with time is observed \cite{Koiter1945}.}
A convergence study is performed for both 2D and 3D simulations: the models are considered at convergence if further mesh refinement corresponds to a relative variation of the critical Cauchy number smaller than $1\%$.
Furthermore, the model with the converged refinement is validated against experimental results in the cases corresponding to $w_c/l=1.6,\, b/b_c=0.46$ and $w_c/l=0.6,\, b/b_c=0.8$, with an agreement in terms of buckling threshold within $1\%$.

\begin{figure}[!h]
\begin{tikzpicture}[scale=1]

  \node[inner sep=0pt,black] at (-6,1.85)
	{a)};
   \node[inner sep=0pt,black] at (3,1.85)
	{b)};
 
\draw[thick] (-0.5-5,0) rectangle (6.5-5,1.5);

  \node[inner sep=0pt,black] at (-2,1.2)
	{no slip};
  \node[inner sep=0pt,black] at (-2,0.3)
	{no slip};
   \node[inner sep=0pt,black] at (-0.5,0.75)
	{clamp};
 
  \node[inner sep=0pt,black,rotate=90] at (-5.7,0.75)
	{inlet};
  \node[inner sep=0pt,black,rotate=90] at (1.7,0.75)
	{outlet};
 
\draw[thick,fill=green] (2-5,0.75-0.05) rectangle (4-5,0.75+0.05);

\node[inner sep=0pt,black] at (7,0.75)
{
\begin{tikzpicture}
 \begin{scope}[x={(4cm,0cm)},y={({cos(30)*2.2cm},{sin(30)*2.2cm})},
    z={({cos(90)*0.8cm},{sin(90)*0.8cm})},line join=round,thick,fill opacity=0.2]
  \draw[fill=none, black!50!white] (0,0,0) -- (0,0,1) -- (0,1,1) -- (0,1,0);
  \draw[fill=blue, black!50!white] (0,0,0) -- (0.5,0,0) -- (0.5,1,0) -- (0,1,0) -- cycle;
  \draw[fill=none, black!50!white] (0,1,0) -- (0.5,1,0) -- (0.5,1,1) -- (0,1,1) -- cycle;
  \draw[fill=none] (0.5,0,0) -- (0.5,0,1) -- (0.5,1,1) -- (0.5,1,0) -- cycle;
  \draw[fill=none] (0,0,1) -- (0.5,0,1) -- (0.5,1,1) -- (0,1,1) -- cycle;
  \draw[fill=none] (0,0,0) -- (0.5,0,0) -- (0.5,0,1) -- (0,0,1) -- cycle;
 \end{scope}

\end{tikzpicture}
};

\node[inner sep=0pt,black] at (7.1,0.6)
{
\begin{tikzpicture}

  \begin{scope}[x={(0.6cm,0cm)},y={({cos(30)*1.1cm},{sin(30)*1.1cm})},
    z={({cos(90)*0.3cm},{sin(90)*0.3cm})},line join=round,thick,fill opacity=0.6,black]
  \draw[fill=mygreen] (0,0,0) -- (0,0,1) -- (0,1,1) -- (0,1,0);
  \draw[fill=mygreen] (0,0,0) -- (0.5,0,0) -- (0.5,1,0) -- (0,1,0) -- cycle;
  \draw[fill=mygreen] (0,1,0) -- (0.5,1,0) -- (0.5,1,1) -- (0,1,1) -- cycle;
  \draw[fill=mygreen] (0.5,0,0) -- (0.5,0,1) -- (0.5,1,1) -- (0.5,1,0) -- cycle;
  \draw[fill=mygreen] (0,0,1) -- (0.5,0,1) -- (0.5,1,1) -- (0,1,1) -- cycle;
  \draw[fill=mygreen] (0,0,0) -- (0.5,0,0) -- (0.5,0,1) -- (0,0,1) -- cycle;
 \end{scope}
 
\end{tikzpicture}
};

   \node[inner sep=0pt,black] at (9.1,0.5)
	{clamp};
 \draw[-] (8.65,0.5) -- (7.55,0.85);
 
    \node[inner sep=0pt,black] at (9.2,0)
	{symmetry plane};
 \draw[-] (8,0) -- (7.5,0.4);

     \node[inner sep=0pt,black] at (5.5,0.4)
	{inlet};
      \node[inner sep=0pt,black] at (7.5,1.4)
	{outlet};

                 \draw[->,black,thick] (4.1,0.3) -- (4.1,0.8);
                 \draw[->,black,thick] (4.1,0.3) -- (3.7,0.3);
                 \draw[->,black,thick] (4.1,0.3) -- (4.5,0.65);                  
                 \node[inner sep=0pt] at (3.7,0.09)
	{$y$};
                 \node[inner sep=0pt] at (4.6,0.5)
	{$x$};	
                 \node[inner sep=0pt] at (3.9,0.9)
	{$z$};	
 
\end{tikzpicture}
\caption{(a) 2D domain depicting the different boundary conditions. (b) 3D domain (not to scale) with the symmetry plane in grey and the different boundary conditions.}
\label{fig:comsol}
\end{figure}
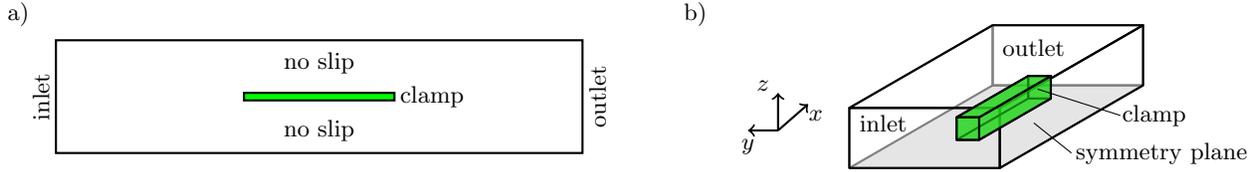

Figure \ref{fig:comsol} (a) shows the 2D geometry of the beam (green) immersed in a rectangular channel. The vertical edge on the left is the inlet, where a parabolic velocity profile is given as a boundary condition. The right vertical edge is the outlet, where the pressure is set to $0$. All other edges are assigned a no-slip boundary condition and Stokes equations are solved within the channel. The beam is modeled as a Hookean solid undergoing small strains but large displacement gradients. The vertical right edge of the beam is the clamp, where the displacement vector is set to zero.

Figure \ref{fig:comsol} (b) shows the 3D geometry (not to scale) where the boundary conditions are applied similarly to the 2D case. The only difference in this case is the introduction of a symmetry plane to reduce the computational cost by taking advantage of the symmetry with respect to the xy-plane. 

For both 2D and 3D simulations, a parametric study is performed to identify the minimum length of the numerical channel ($L\simeq10l$), above which results become invariant upon further changes in the length.

\section{Buckling instability due to transversal pressure loads}

\subsection{Pressure load due to small deflections}
\begin{figure}
\centering
\includegraphics[scale=0.45]{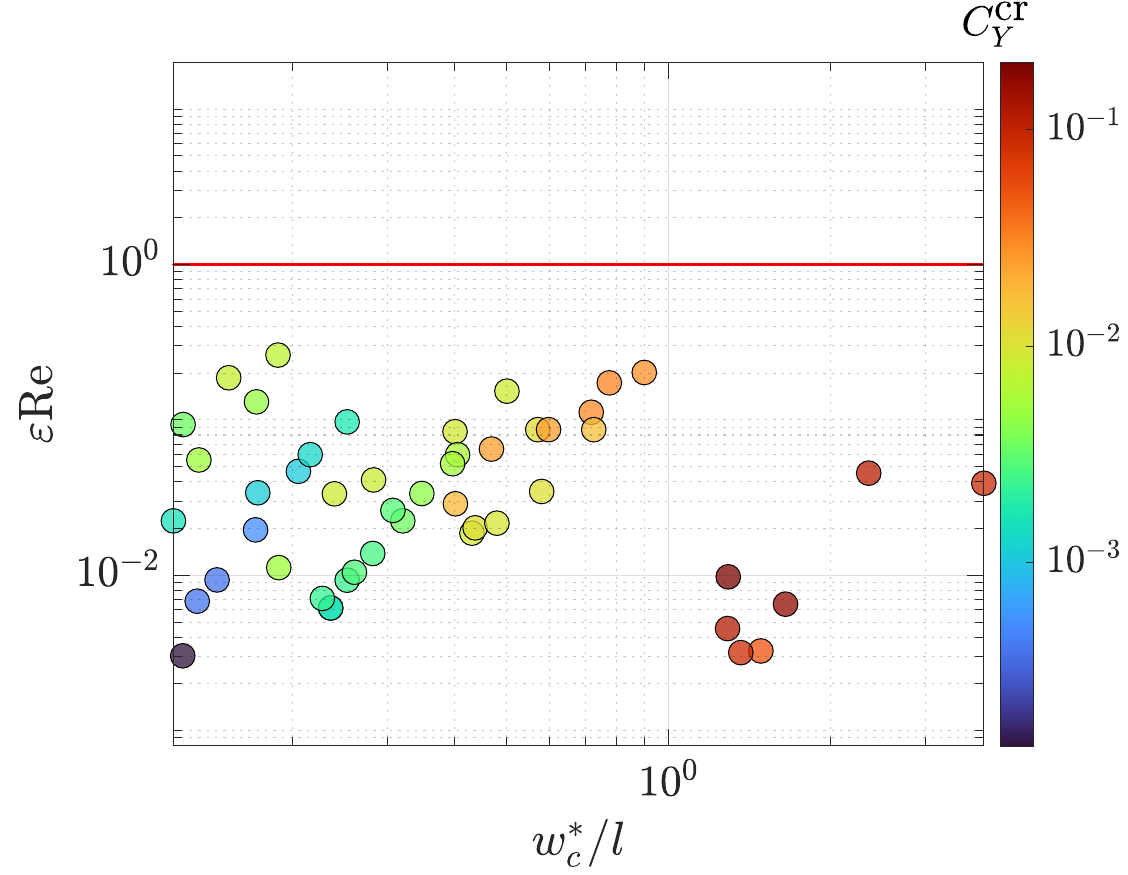}
\caption{{Values of $\varepsilon Re$ versus $w_c^*/l$, for all experiments. The red line corresponds to $\varepsilon Re = 1$, which is the limit beyond which the asymptotic expansion theoretically breaks down.}}
\label{fig:Reynolds}
\end{figure}

We consider the flow of a viscous fluid of viscosity $\mu$ occurring in the {three}-dimensional {rectangular channel} of {width}  $H=(w_\textup{c}-h)/2$ between the straight beam and the upper and lower walls of the channel {and height $b_c$}. As shown in Figure \ref{fig:comsol}, we introduce the reference frame $(x,y,z)$ aligned with the beam (from the free-edge to the clamp), the width and the height of the channel, respectively, with the origin located at the centroid of the free section of the beam. The {Navier-}Stokes equations governing the motion, rendered non-dimensional with the length of the beam, the inlet maximum velocity and the characteristic pressure $\mu U_\textup{max}/l$, read
\begin{equation}
 \nabla \cdot \bar{\boldsymbol{u}}=0\,, \quad {{\mathrm{Re}  \bar{\boldsymbol{u}}} \nabla  \bar{\boldsymbol{u}} }=-\nabla \bar{p} +  \nabla^2  \bar{\boldsymbol{u}}\,,
\end{equation}
where {$\mathrm{Re}=U_\textup{max}l/\nu$ is the Reynolds number, and}  $\bar{p}$ and $\bar{\boldsymbol{u}}$ are the non-dimensional pressure and velocity field, respectively.
This equation is coupled with the no-slip conditions at $y=\pm H/2$ {and $z=\pm b_c/2$}. 
As observed in Figure~1 of the main text, when the beam is long enough, the pressure does not vary appreciably along the $y$ direction.
This result can be derived from the lubrication approximation here employed. Under the assumption $H{,b_c} \ll l$, gradients along the $y$ direction  are much larger than those along the $x$ direction, i.e. $\partial_x \ll \partial_y{,\partial_z}$ \citep{duprat2016fluid}. 
The following multiple scale expansion is thus employed:
\begin{equation}
    \partial_{\bar{x}} =\varepsilon \partial_X\,, \quad \partial_{\bar{y}} = \partial_Y\,{,\quad \partial_{\bar{z}} = \partial_Z\,},\,\, \varepsilon = {D_h^*}/ l  \ll 1\,{,}
    \end{equation}
    {where $D_h^*=w_c b_c/(b_c+w_c/2)$ is the hydraulic diameter to account for the rectangular section of each gap between the walls and the beam, neglecting its small thickness.}
The continuity equation reads    
\begin{equation}
 {\varepsilon} \partial_X \bar{u}_x+ \partial_Y \bar{u}_y {+ \partial_Z \bar{u}_z}=0 \rightarrow \bar{u}_y, {\bar{u}_z} \sim \varepsilon \bar{u}_x\,,
\end{equation}
thus implying that the $y$- {and $z$-}component{s} of the velocity field {are} of order $\varepsilon$ when compared to the $x$-component.
We now expand the velocity field 
\begin{equation}
    \bar{p}=\bar{p}^{(0)}+\mathcal{O}(\varepsilon)\,, \quad \bar{u}_x=\bar{u}_x^{(0)}+\mathcal{O}(\varepsilon)\,,\quad \bar{u}_y=\varepsilon \bar{u}_y^{(1)}+\mathcal{O}(\varepsilon^2)\,{,\quad \bar{u}_z=\varepsilon \bar{u}_z^{(1)}+\mathcal{O}(\varepsilon^2)}\,,
\end{equation}
so that the asymptotic expansion of the {Navier-}Stokes equations become
{
\begin{equation}
   {\varepsilon \mathrm{Re} \left(\bar{u}_x^{(0)}  \partial_X \bar{u}_x^{(0)} +\bar{u}_y^{(0)}  \partial_Y \bar{u}_x^{(0)} +\bar{u}_z^{(0)} \partial_Z \bar{u}_x^{(0)}  +\mathcal{O}(\varepsilon) \right)} =  - \varepsilon \partial_X \bar{p}^{(0)} +\mathcal{O}(\varepsilon)+  \left( \varepsilon^2 \partial_X^2 + \partial_Y^2 + \partial_Z^2\right) \left(\bar{u}_x^{(0)}+\mathcal{O}(\varepsilon)\right)\,,
      \end{equation}
   \begin{equation*} 
   \mathcal{O}(\varepsilon^2) = -\partial_Y \bar{p}^{(0)}+\mathcal{O}(\varepsilon)+ \varepsilon  \left( \varepsilon^2 \partial_X^2 + \partial_Y^2+ \partial_Z^2\right) \left(\bar{u}_y^{(1)}+\mathcal{O}(\varepsilon)\right)\,,
\end{equation*}
   \begin{equation*} 
   \mathcal{O}(\varepsilon^2) = -\partial_Z \bar{p}^{(0)}+\mathcal{O}(\varepsilon)+ \varepsilon  \left( \varepsilon^2 \partial_X^2 + \partial_Y^2 + \partial_Z^2\right) \left(\bar{u}_z^{(1)}+\mathcal{O}(\varepsilon)\right)\,.
\end{equation*}
The convective term on the LHS of the x-component of the Navier-Stokes equations is negligible as long as $\varepsilon \mathrm{Re} \ll 1$. We thus define the limit value below which this convective term can be safely neglected as $\mathrm{Re}_{c}=1/\varepsilon$. Within our experiments, the value of $\varepsilon \mathrm{Re}$ remains at least one order of magnitude smaller than unity, as shown in Figure \ref{fig:Reynolds}, thus ensuring that the convective terms are negligible within this framework and that the Poiseuille flow approximation presented in the following can be safely employed.}

\begin{figure}
\centering
\begin{tikzpicture}
    
  \node[inner sep=0pt,black] at (-6.5,2.2)
	{a)};

         \node[inner sep=0pt] at (1.7,0.09)
	{ \includegraphics[width=0.9\textwidth]{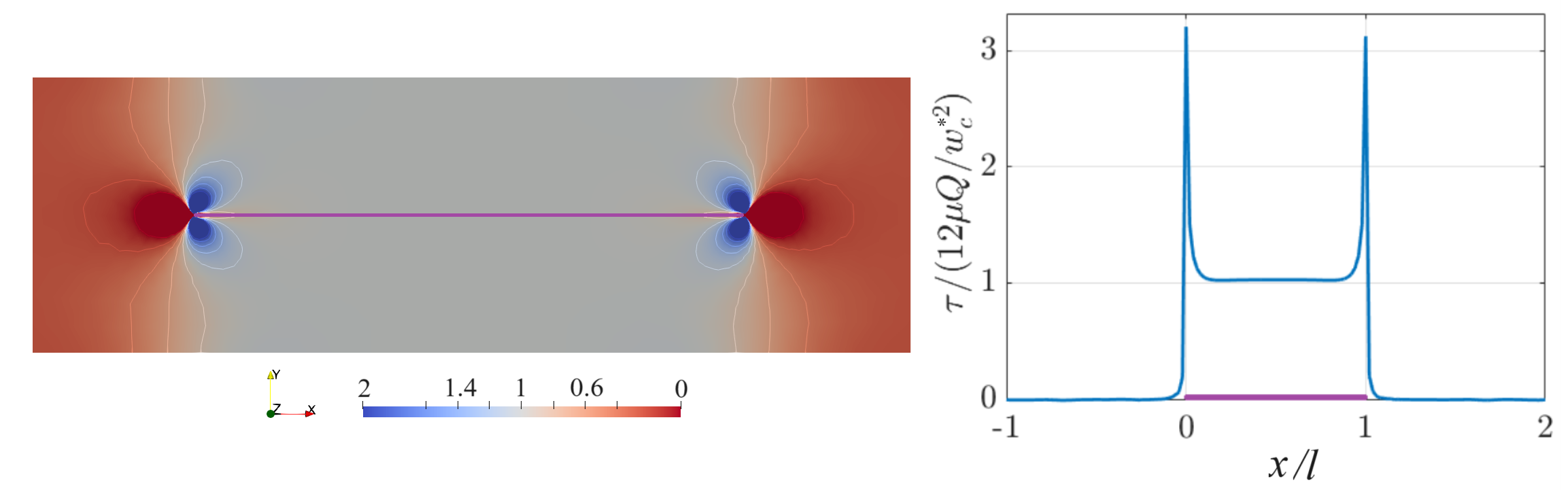} };

\draw [fill=white,white] (-3,-1.75) rectangle (-3.7,-1.2);
   \node[inner sep=0pt,black] at (-4.85,-1.7)
	{$x$};
    \draw[->,black,thick] (-5.5,-1.69) -- (-5,-1.69);

       \node[inner sep=0pt,black] at (-5.3,-1.2)
	{$y$};
    \draw[->,black,thick] (-5.49,-1.7) -- (-5.49,-1.2);

   \node[inner sep=0pt,black] at (3.1,2.2)
	{b)};   
  \end{tikzpicture}
 \caption{Results of two-dimensional simulations below the buckling threshold. (a) Spatial distribution of $G/\left(\frac{6 \mu Q}{(w_c^*/2)^3}\right)$, which exhibits a constant unitary value in the regions above and below the beam. (b) Variation with $x$ of the rescaled wall shear stress $\tau/(12\mu Q / w_c^{*2})$ along a horizontal line lying on the upper surface of the beam. Except for two peaks at the beam tips, the value is constant and equal to one along the beam (purple line).}
    \label{fig:2dnum}
\end{figure}

{We now focus on a two-dimensional flow (i.e., $\partial_Z=0$), and we employ the classical assumption of a Poiseuille flow driven by a constant pressure gradient $\varepsilon \partial_X \bar{p}^{(0)} =\mathcal{O}(1)$. The flow equations become:
}

\begin{equation}
   \varepsilon \partial_X \bar{p}^{(0)} +\mathcal{O}(\varepsilon)=  \left( \varepsilon^2 \partial_X^2 + \partial_Y^2\right) \left(\bar{u}_x^{(0)}+\mathcal{O}(\varepsilon)\right)\,, \quad   \partial_Y \bar{p}^{(0)}+\mathcal{O}(\varepsilon)= \varepsilon  \left( \varepsilon^2 \partial_X^2 + \partial_Y^2\right) \left(\bar{u}_y^{(1)}+\mathcal{O}(\varepsilon)\right)\,.
\end{equation}
At leading order, the {Navier-}Stokes equations simplify to
\begin{equation}
   \varepsilon \partial_X \bar{p}^{(0)}= 
     \partial_Y^2 \bar{u}_x^{(0)}, \quad   \partial_Y \bar{p}^{(0)}= 0\,,
\end{equation}
i.e., the pressure $p(x)$ does not vary along the $y$ direction, at leading order.
Upon definition of the constant pressure gradient $\partial_x p:=-G$ and integration along the $y$ direction with no-slip conditions at $y=\pm H/2$, reverting to dimensional, physical, variables, and dropping leading order notation for the sake of simplicity, one obtains
\begin{equation}
    p(x)=p(x=0)- G x\,, \quad u_x(y)=\frac{G H^2}{8\mu}\left( 1- (2y/H)^2 \right)\,.
\end{equation}
The constant pressure gradient $G$ that ensures a constant flow rate $Q/2$ at each section for a streamwise-invariant flow (half of the total one, equally divided between upper and lower sides of the beam) reads
\begin{equation}
    G= \frac{6 \mu Q}{H^3}\,, 
\end{equation}
which shows a very good agreement with the spatial distribution of $G$ in two-dimensional numerical simulations when $H=(w_c-h)/2=w_c^*/2$, as depicted in Figure \ref{fig:2dnum} (a).

When dealing with very small deflections of the beam along the $y$-direction $w(x) \ll l$, i.e. $H=w_c^*/2 \pm w(x)$, this framework is still assumed valid, i.e. 
\begin{equation}
    G= \frac{6 \mu Q}{(w_c^*/2\pm w(x))^3}\,, \quad p(x)=p(x=0)- G x\,, 
\end{equation}
where the sign $\pm$ depends on the side of the channel, with the negative sign for $y>0$ and vice-versa.
Neglecting edge effects in the upstream leading edge of the beam, the pressure on both sides reads:
\begin{equation}
    \begin{aligned}
& p_{+}(x) =- 6\mu Q x /\left(w_c^* / 2-w(x)\right)^3 +p(x=0)\,, \\
& p_{-}(x) =- 6\mu Q x /\left(w_c^* / 2+w(x)\right)^3 +p(x=0)\,.
\end{aligned}
\end{equation}
At a fixed downstream position, the pressure difference between the upper (+) and lower (-) part of the beam is expressed as a Taylor series for $w(x)/w_c^* \rightarrow 0$:
\begin{equation}
\begin{aligned}
    \Delta p (x)\left(\boldsymbol{e}_y\right)=&-p_{+}\left(\boldsymbol{e}_y\right)+p_{-}\left(\boldsymbol{e}_y\right) = \frac{576 \mu Q x}{w_c^{*3}} \left(\frac{w(x)}{w_c^*}\right)\boldsymbol{e}_y +\mathcal{O}(w^3(x)) \approx \alpha  x {w(x)}\boldsymbol{e}_y\,,
\end{aligned}   
\end{equation}
where $\boldsymbol{e}_y$ is the unit base vector along $y$ {and $\alpha$ is defined as a result}. This force acts along the same direction of the displacement and is analogous to a Winkler foundation with a negative spring stiffness $\alpha x$, and can also be seen as a fluid compliance due to pressure.

\subsection{Instability threshold due to pressure load}
The transverse load per unit transversal length thus reads
\begin{equation}
    q_y= \frac{576 \mu Q x}{w_c^{*4}} w(x)\,,
\end{equation}
which is included in the linear beam equation \cite{Winkler1867}
\begin{equation}
E \hat{I}w^{\prime \prime \prime \prime}(x)- \frac{576 \mu Q x}{w_c^{*4}} w(x)=0\,.
\end{equation}
Upon non-dimensionalization with the length of the beam $l$ and introduction of the Cauchy number $C_Y=\frac{\mu U_\textup{max} l^2}{E\hat{I}}$ (at the inlet, $Q=(2/3)U_\textup{max}w_c$), the equation reads\begin{equation}
    \bar{w}^{\prime \prime \prime \prime}(\bar{x})-384 C_Y\left(\frac{l}{w_c^{*}}\right)^3 \bar{x} \bar{w}(\bar{x})=0 \rightarrow \bar{w}^{\prime \prime \prime \prime}(\bar{x})-\bar{\alpha} \bar{x} \bar{w}(\bar{x})=0\,.
    \label{eq:buckling_p}
\end{equation}
Note that we assumed $w_c^*/w_c \approx 1 $ in our calculations. Indeed, $w_c^*/w_c\in[0.7,1]$ in our experiments and simulations, thus not affecting the scaling in an appreciable manner.
The general solution of this equation is written via hypergeometric functions
\begin{multline}
    \bar{w}(\bar{x})=\frac{(-1)^{3/5} c_4 \bar{\alpha}^{3/5} \bar{x}^3 \, _0F_3\left(;\frac{6}{5},\frac{7}{5},\frac{8}{5};\frac{\bar{\alpha} \bar{x}^5}{625}\right)}{25\ 5^{2/5}}+\frac{(-1)^{2/5} c_3 \bar{\alpha}^{2/5} \bar{x}^2 \, _0F_3\left(;\frac{4}{5},\frac{6}{5},\frac{7}{5};\frac{\bar{\alpha} \bar{x}^5}{625}\right)}{5\ 5^{3/5}} +\\+\frac{\sqrt[5]{-1} c_2 \sqrt[5]{\bar{\alpha}} \bar{x} \, _0F_3\left(;\frac{3}{5},\frac{4}{5},\frac{6}{5};\frac{\bar{\alpha} \bar{x}^5}{625}\right)}{5^{4/5}}  +c_1 \, _0F_3\left(;\frac{2}{5},\frac{3}{5},\frac{4}{5};\frac{\bar{\alpha} \bar{x}^5}{625}\right)\,,
    \label{eq:sol_hyp}
\end{multline}
with the classical free-edge ($\bar{w}''(0)=\bar{w}'''(0)=0$) and clamp conditions ($\bar{w}(1)=\bar{w}'(1)=0$). Non-trivial solutions of this problem are found by imposing a zero determinant for the system matrix of equations stemming from the boundary conditions, leading to 
\begin{multline}
f(\bar{\alpha})=3 \bar{\alpha} \, _0F_3\left(;\frac{3}{5},\frac{4}{5},\frac{6}{5};\frac{\bar{\alpha}}{625}\right) \, _0F_3\left(;\frac{7}{5},\frac{8}{5},\frac{9}{5};\frac{\bar{\alpha}}{625}\right)-\,\\- _0F_3\left(;\frac{2}{5},\frac{3}{5},\frac{4}{5};\frac{\bar{\alpha}}{625}\right) \left(72 \, _0F_3\left(;\frac{3}{5},\frac{4}{5},\frac{6}{5};\frac{\bar{\alpha}}{625}\right)+\bar{\alpha} \, _0F_3\left(;\frac{8}{5},\frac{9}{5},\frac{11}{5};\frac{\bar{\alpha}}{625}\right)\right)=0\,. 
\end{multline}
A simple approximation of this expression is found by exploiting the Taylor series:
\begin{equation}
    f(\bar{\alpha})=-\frac{11 \bar{\alpha}^2}{12600}+\frac{6 \bar{\alpha}}{5}-72+\mathcal{O}(\bar{\alpha}^3) \rightarrow \bar{\alpha}_1=7560/11 - (720 \sqrt{91})/11 \approx 62.88\,,
\end{equation}
very close to the numerical value $62.8531$ from the exact $f(\bar{\alpha})$. 
\begin{figure}
    \centering
    \includegraphics[scale=0.85]{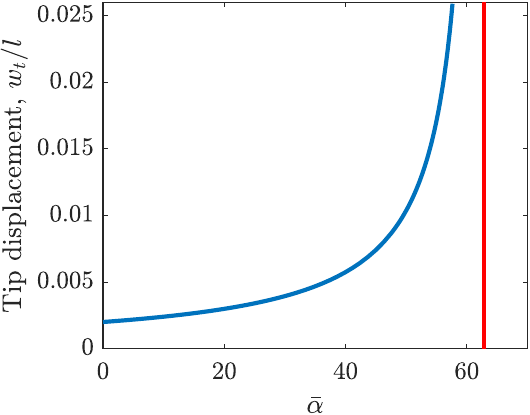}
    \caption{Tip displacement in the presence of imperfections for $\bar{w}_0=0.001$ and $\bar{w}_1=\bar{w}_2=0$. The red vertical line denotes the analytical instability threshold.}
    \label{fig:tip}
\end{figure}
Therefore, the following expression for the critical Cauchy number for the instability as a function of the gap-to-beam length ratio is obtained:
\begin{equation}
    384 C_Y^{\textup{cr}}\left(\frac{l}{w_c^*}\right)^3=\bar{\alpha}_1 \rightarrow C_Y^{\textup{cr}} \approx 0.1637\left(\frac{w_c^*}{l}\right)^3\,. 
\end{equation}

\subsection{The effect of imperfections of the beam position in the tip displacement at buckling}

Small imperfections can be modeled by modifying the gap as $H=w_c^*/2\pm w(x)+w_0+w_1 x + w_2x^{2} + ...\,$. Here, $w_0$ represents an offset in the position of the beam with respect to the centerline of the channel, $w_1$ represents a small rotation with respect to $\boldsymbol{e}_x$, and $2w_2$ represents a linear natural curvature.
Truncating  the imperfection at order $\mathcal{O}(x^2)$, a simple solution based on the previous one can be obtained. This solution gives a flavour on the observable effects of imperfections in experiments. We introduce the variable transformation
\begin{equation}
    \bar{w}^*(\bar{x})=\bar{w}(\bar{x})+\bar{w}_0+\bar{w}_1 \bar{x}+\bar{w}_2 \bar{x}^{2}\,.
\end{equation}
Equation \eqref{eq:buckling_p} together with its boundary conditions can be re-written as
\begin{equation}
{{\bar{w}}^*}{''''}(\bar{x})-\bar{\alpha} \bar{x} {\bar{w}}^{*}(\bar{x})=0\,, \quad {\bar{w}}^{*}(1)={\bar{w}}_0+{\bar{w}}_1+{\bar{w}}_2\,, \,{{\bar{w}}^{*}}{'}(1)=-{\bar{w}}_1+2{\bar{w}}_2\,, \,{{\bar{w}}^{*}}{''}(0)=2{\bar{w}}_2\,, \,{{\bar{w}}^*}{'''}(0)=0\,.
\label{eq:imperfections}
\end{equation}
The analytical solution is formally analogous to Eq. \eqref{eq:sol_hyp} where constants satisfy the boundary conditions. The numerical tip displacement $\bar{w}(0)=w_t/l$ obtained for different values of $\bar{\alpha}$ from Eq. \eqref{eq:imperfections} for $\bar{w}_0=0.001$, $\bar{w}_1=\bar{w}_2=0$, is reported in Figure \ref{fig:tip}. Through a Taylor expansion, we can approximate the tip displacement as follows:
\begin{equation}
    w_t/l=\bar{\alpha}\frac{ (-84 \bar{w}_0-35 \bar{w}_1-18 \bar{w}_2)}{2520}+\bar{\alpha}^2\frac{ (-10923 \bar{w}_0-4690 \bar{w}_1-2464 \bar{w}_2)}{19958400}+\mathcal{O}(\bar{\alpha}^3)\,,
\end{equation}
i.e. the tip displacement is initially linear with the flow rate (see Figure \ref{fig:tip}), as observed in the experiments, with a progressive divergence when reaching the asymptotic value given by the instability threshold.

\begin{figure}[h]
\includegraphics[width=0.35\textwidth]{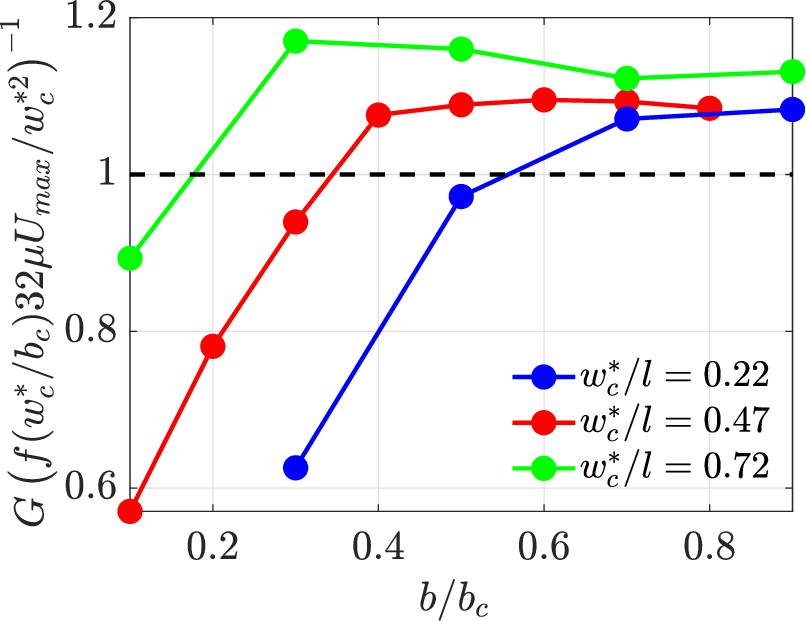}
\includegraphics[width=0.35\textwidth]{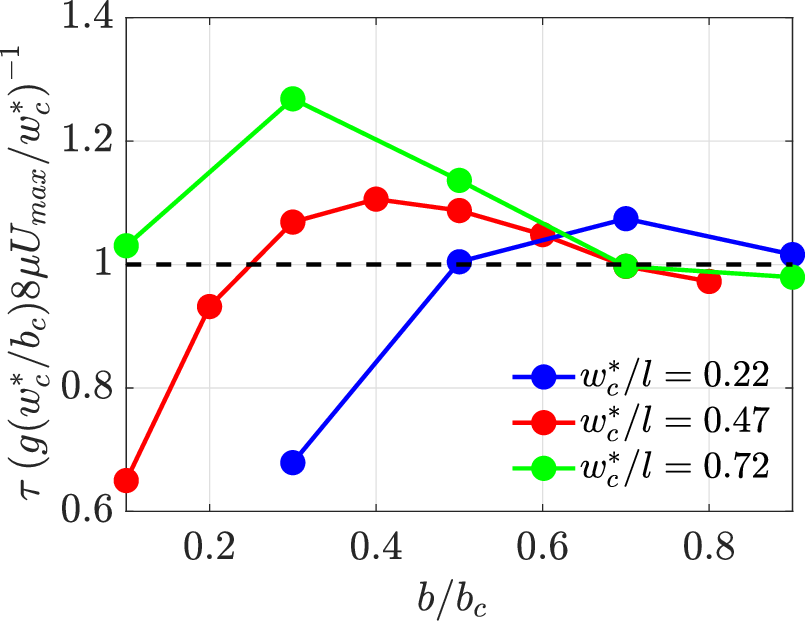}
\caption{{Surface-average on the upper wall of the beam (obtained from numerical simulations) of (a) pressure gradient and (b) wall shear stresses, rescaled with the theoretical values as functions of $b/b_c$. }}
\label{fig:3d_num}
\end{figure}

{\section{Three-dimensional effects: comparison between theory and numerics}}
{
The analytical values of $f(w^*_c/b_c)$ and $g(w^*_c/b_c)$ as functions of $w^*_c/b_c$ are reported in Figure 3 of the main text. For $w^*_c/b_c \rightarrow 0$, these functions approach the unity, i.e. the values of $G$ and $\tau$ are well approximated by their two-dimensional counterparts. Conversely, these values increase when the channel becomes narrow. These theoretical values well agree with those extracted from numerical simulations, obtained by averaging quantities on the upper wall of the beam, with varying $b/b_c$ and $w^*_c/l$, as shown in Figure \ref{fig:3d_num}. Small deviations are imputed to local distributions due to edge effects as well as integral approximations, and do not alter the qualitative and quantitative agreement of the scaling.
}

\newpage

\setcounter{figure}{0}
\makeatletter
\renewcommand{\figurename}{MOV.}

\pagebreak

\section{Supplementary movies}

\begin{figure}[!h]
    \centering
    \includegraphics[width=0.345\textwidth]{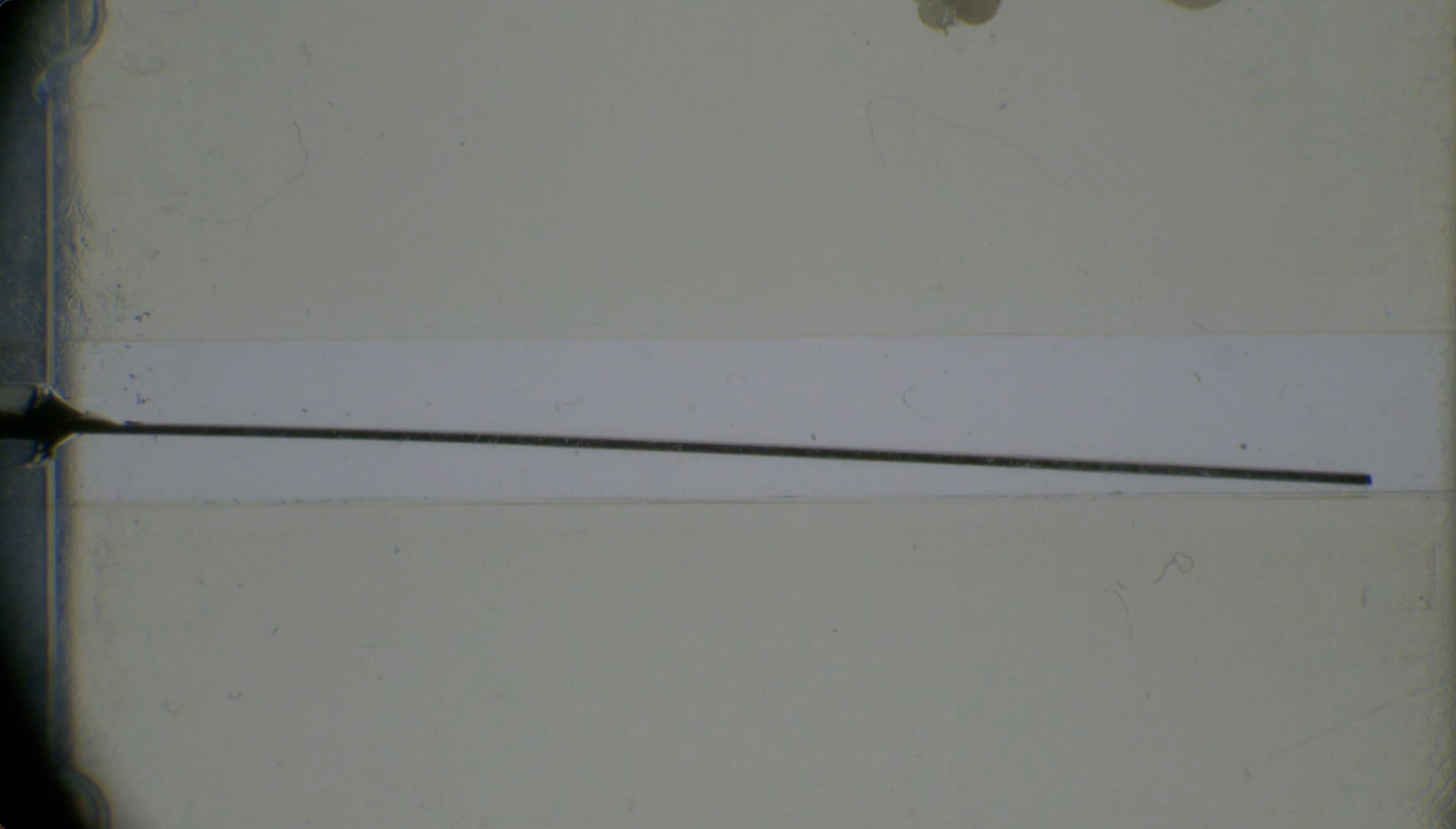}
    \caption{Experiment with $w_c/l=0.13$, $h/l=0.0066$, $b/b_c=0.46$ at $Q=50$ mL min$^{-1}$, higher than the buckling threshold. The beam, made of PET (Mylar\textregistered), can be seen in black, while touching the wall of the channel.}
    \label{fig:beam34_Movie}
\end{figure}

\begin{figure}[!h]
    \centering
    \includegraphics[width=0.345\textwidth]{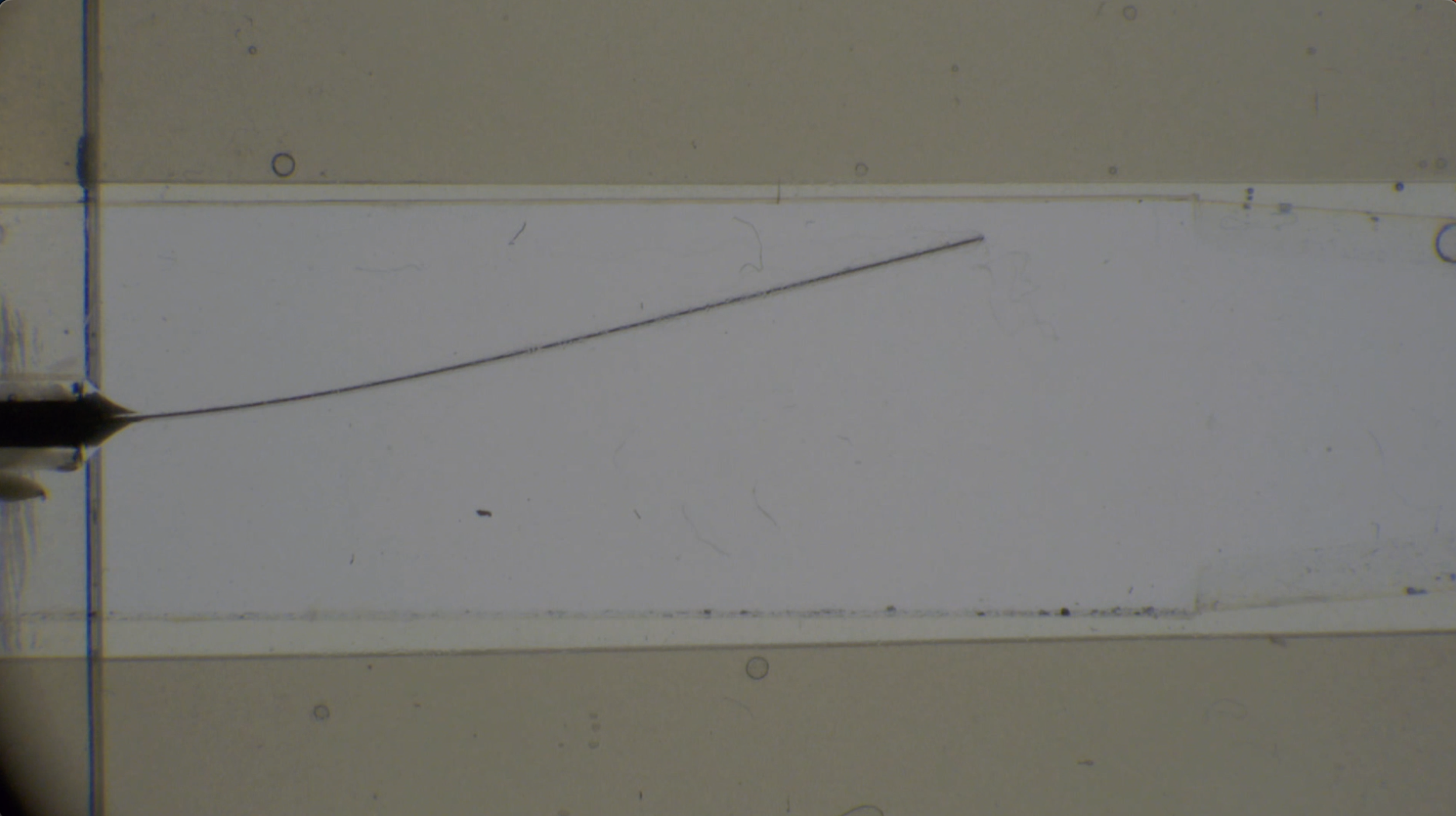}
    \caption{Experiment with $w_c/l=0.47$, $h/l=0.00354$, $b/b_c=0.46$ at $Q=35$ mL min$^{-1}$, higher than the buckling threshold. The beam is made of PET (Mylar\textregistered).}
    \label{fig:beam42_Movie}
\end{figure}

\begin{figure}[!h]
    \centering
    \includegraphics[width=0.345\textwidth]{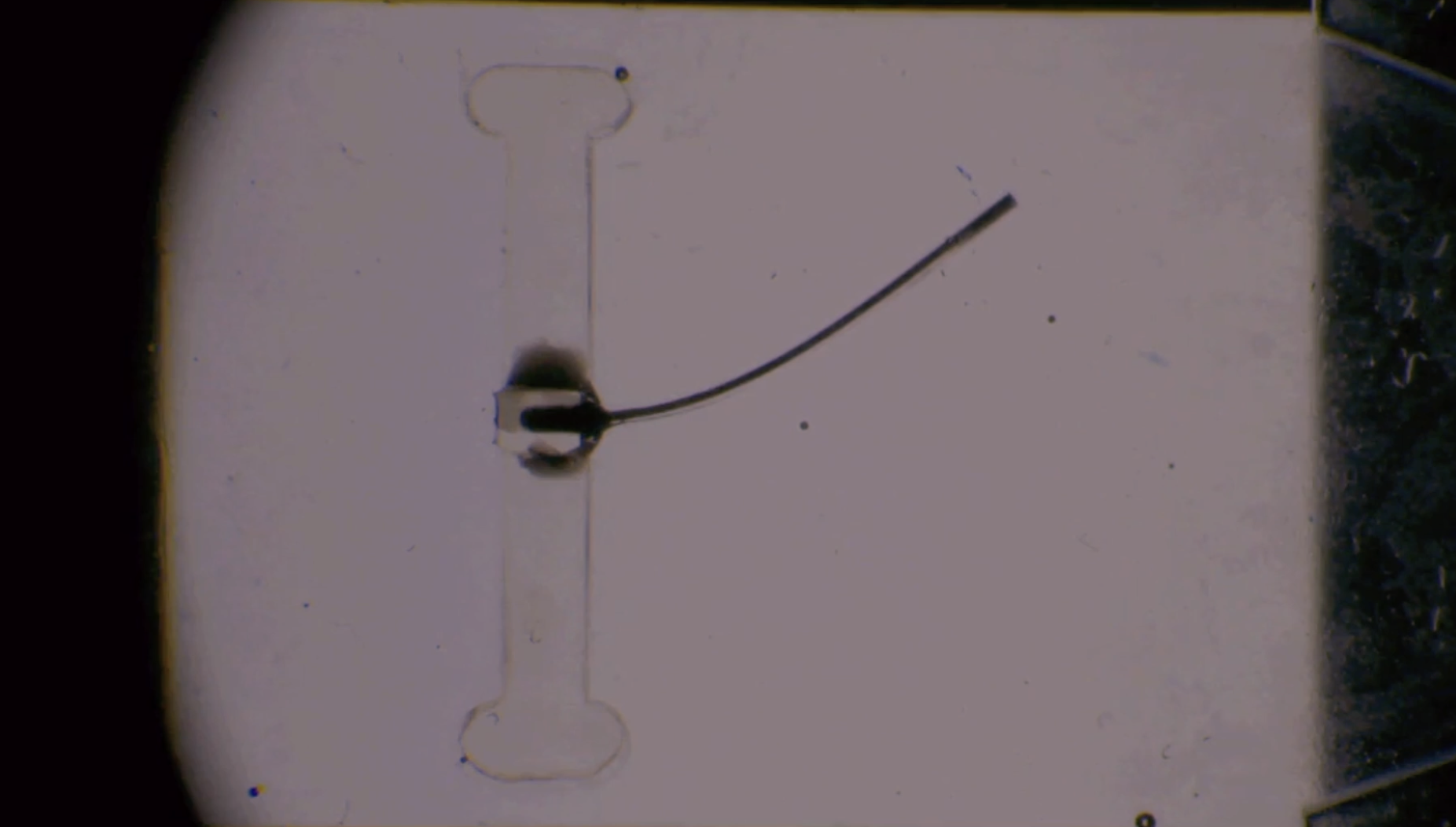}
    \caption{Experiment with $w_c/l=1.67$, $h/l=0.0175$, $b/b_c=0.46$ at $Q=16$ mL min$^{-1}$, higher than the buckling threshold. The beam is made of VPS-32.}
 
    \label{fig:beam7_Movie}
\end{figure}

\end{document}